\documentclass[aps,twocolumn,showpacs,email,amssymb,nobibnotes,pre]{revtex4}
\usepackage{amssymb,amsmath}
\usepackage{xcolor}
\usepackage{graphicx}
\usepackage[caption=false]{subfig}
\usepackage{mathtools}
\usepackage{natbib}
\usepackage{epstopdf}

\newcommand\numberthis{\addtocounter{equation}{1}\tag{\theequation}}

\newcommand{\comments}[1]{}

\def\d{\mathrm{d}}
\newcommand{\p}{\Pi}		

\newcommand{\avg}[1]{\left\langle{#1}\right\rangle}

\newcommand{\beq}{\begin{equation}}
\newcommand{\eeq}{\end{equation}}
\newcommand{\bal}{\begin{aligned}}
\newcommand{\eal}{\end{aligned}}

\newcommand{\be}{\begin{equation}}
\newcommand{\ee}{\end{equation}}
\newcommand{\bd}{\begin{displaymath}}
\newcommand{\ed}{\end{displaymath}}
\newcommand{\BE}{\begin{eqnarray}}
\newcommand{\EE}{\end{eqnarray}}

\begin{document}
\title{Intrinsic noise in systems with switching environments}
\author{Peter G. Hufton}
\email{peter.hufton@postgrad.manchester.ac.uk}
\affiliation{Theoretical Physics, School of Physics and Astronomy,
The University of Manchester, Manchester M13 9PL, United Kingdom}

\author{Yen Ting Lin}
\email{yenting.lin@manchester.ac.uk}
\affiliation{Theoretical Physics, School of Physics and Astronomy,
The University of Manchester, Manchester M13 9PL, United Kingdom}
\author{Tobias Galla}
\email{tobias.galla@manchester.ac.uk}
\affiliation{Theoretical Physics, School of Physics and Astronomy,
The University of Manchester, Manchester M13 9PL, United Kingdom}
\author{Alan J. McKane}
\email{alan.mckane@manchester.ac.uk}
\affiliation{Theoretical Physics, School of Physics and Astronomy,
The University of Manchester, Manchester M13 9PL, United Kingdom}

\begin{abstract}
We study individual-based dynamics in finite populations, subject to randomly switching environmental conditions. These are inspired by models in which genes transition between on and off states, regulating underlying protein dynamics. Similarly switches between environmental states are relevant in bacterial populations and in models of epidemic spread. Existing piecewise-deterministic Markov process (PDMP) approaches focus on the deterministic limit of the population dynamics while retaining the randomness of the switching. Here we go beyond this approximation and explicitly include effects of intrinsic stochasticity at the level of the linear-noise approximation. Specifically we derive the stationary distributions of a number of model systems, in good agreement with simulations. This improves existing approaches which are limited to the regimes of fast and slow switching.
\end{abstract}


\maketitle
\section{Introduction}
\label{sec:introduction}
There is now a broad consensus that noise plays a crucial role in most dynamical systems in biology, chemistry and in the social sciences. The theory with which to describe these stochastic processes is well established and has its roots in statistical physics. Modelling tools such as master equations, Fokker--Planck equations, and Langevin dynamics are standard and can be found in a number of textbooks \cite{van1992stochastic,gardiner1985handbook, risken1984fokker}. Much of this work focuses on processes between discrete interacting individuals, which can be members of a population in epidemiology \cite{Keeling,mode2000stochastic,black2010stochastic}, atoms or molecules in chemical reaction systems \cite{wilkinson2011stochastic,gillespie2007stochastic}, or proteins in the context of gene regulatory networks \cite{kepler2001stochasticity,walczak2012analytic}. Many of these models are Markovian and their natural description is in terms of a master equation which describes the time evolution of the underlying probability distribution of microstates. 

Solving the master equation analytically is often a difficult task: only very simple linear model dynamics allow further treatment \cite{van1992stochastic,gardiner1985handbook, risken1984fokker,shahrezaei2008analytical, kumar2014exact}. It is therefore common to employ approximation techniques, most notably ones built around the assumption that the size of the population is large, but finite. The inverse system size (or its square root) then serves as a small parameter in which an expansion can be performed. The lowest order in this expansion corresponds to the limit of infinite systems, and provides a deterministic description devoid of stochasticity. The second-order term in the expansion introduces some stochastic effects of noise in the population, but approximates the individual-level dynamics by a simpler Gaussian process on a continuum domain \cite{kampen1961power}.
These techniques have been very successful in capturing elements of noise-induced phenomena, for example the weak selection of competing species \cite{lin2012features,lin2015demographicI,lin2015demographicII,constable2015models,PhysRevE.90.042149}, cyclic behaviour, patterns and waves \cite{Biancalani, BiancalaniGallaMcKane, ButlerGoldenfeld}. The main techniques used to characterise these effects are system-size expansion methods, most notably the Kramers--Moyal and van Kampen expansions. The latter is also known as the linear-noise approximation (LNA) \cite{van1992stochastic,gardiner1985handbook}.

These methods are now used routinely for the analysis of individual-based models with intrinsic noise in the weak-noise limit. Most applications so far focus on systems in which the reaction rates are set by constant model parameters, and the only time dependence is in the evolution of the population of individuals itself. Recently, exceptions have gained attention \cite{leibler2010individual, escudero2008persistence, assaf2008population, schuler2005experimental, black2010stochastic, boland2009limit}. In these models, reaction rates vary continuously and deterministically in time, for example to capture periodically varying infection rates to model seasonal variation in epidemic spreading. Crucially, no additional stochasticity is introduced in these dynamics by the environment, and the only discreteness in the dynamics is in the evolution of a finite population of individuals. Other authors have considered models with an environment which varies stochastically and continuously \cite{kamenev2008colored,assaf2013extrinsic,assaf2013cooperation}. 

For many model systems it is more realistic to assume that model parameters switch between different discrete states. This includes phases of antibiotic treatment in the context of bacteria \cite{kussell2005bacterial, kussell2005phenotypic}, genetic switches \cite{kepler2001stochasticity, thattai2001intrinsic, swain2002intrinsic, hayot2004linear, bobrowski2007asymptotic, sherman2014computational, thomas2014phenotypic, duncan2015noise}, evolutionary game theory \cite{ashcroft2014fixation} and predator-prey models in switching environmental conditions \cite{luo2007stochastic, zhu2009competitive}. Such models describe two types of discreteness: that of the state of the environment and that of the population of interacting individuals. In principle, the environmental switching can occur stochastically or follow a deterministic pattern (e.g., prescribed periods of antibiotic treatment, regularly interspersed with periods of no antibiosis). 

In the present work we focus on stochastically switching environments. Assuming again a large, but finite population, the demographic noise in the population can be approximated using the above expansion techniques. The noise relating to switches in the environmental state, however, cannot be dealt with in this way: there is no large parameter to expand in when the number of environmental states is small. 

In models with switching environments the lowest-order expansion in the strength of the intrinsic noise leads to a so-called `piecewise-deterministic Markov process' (PDMP) \cite{davis1984piecewise, faggionato2009non}. In this approximation the dynamics of the population of individuals is described by deterministic rate equations between stochastic switches of the environment. This approach neglects all intrinsic stochasticity from the reaction dynamics {\em within} the population---the population scale is taken to be infinite. The only type of randomness retained is that of the switching of the environmental states. The application of PDMPs has recently gained attention in the description of genetic networks \cite{zeiser2008simulation, zeiser2010autocatalytic, thattai2001intrinsic}.

The PDMP approximation has been surprisingly effective in modeling systems with very large populations \cite{thattai2001intrinsic, zeiser2010autocatalytic, realpe2012demographic}; however, it does not produce accurate results outside this limit. Recently, an alternative approach has incorporated some effects of demographic noise, but it is only valid if there is a very large separation between the time scales of environmental switching and that of the population dynamics \cite{thomas2014phenotypic, duncan2015noise}. Here, we develop the theory further and construct a systematic expansion in the noise strength about the PDMP.

The remainder of this paper is organised as follows: In Section~\ref{sec:simple} we give a more formal introduction to the problem using a relatively simple linear model. We show how the system-size expansion can be applied in the presence of switching environments and we analytically derive the resulting stationary distribution for the linear model. In Section~\ref{sec:general} we construct a more general theory and describe how the method can be applied to a larger set of model dynamics. Section~\ref{sec:examples} contains applications to a number of model systems with nonlinear reaction rates and/or other additional features. In Sec.~\ref{sec:conclusion} we summarise out findings and give an outlook on future work. The Appendix contains further details of the relevant calculations.

\section{Introductory example}
\label{sec:simple}
\subsection{Model definition}
We first focus on a simple example with linear reaction rates. We consider a population of individuals of type $\mathcal{A}$, and we write $n$ for the number of individuals in the population at any given time. Individuals can be created and they can decay, so that the model describes a birth-death process. The death rate per individual is assumed to be a constant $d$. The creation rate is taken to depend on the state of the external environment; individuals are born with rate $\Omega b_\sigma$, where $\sigma$ represents the state of the environment. This can be summarised as follows:
\begin{align*}
\varnothing &\xrightarrow{\mathmakebox[0.5cm]{\Omega b_\sigma}} \mathcal{A}, \\
\mathcal{A} &\xrightarrow{\mathmakebox[0.5cm]{d}} \varnothing, 
\numberthis
\label{eq:simple_reactions_system}
\end{align*}
for times at which the environment is in state $\sigma$. The parameter $\Omega$ has been introduced as per normal convention to set a typical scale of the population size \cite{van1992stochastic,gardiner1985handbook}. If the environment were to be fixed at $\sigma$ the number of individuals in the system would fluctuate around the value $\Omega b_\sigma/d$ in the long run. These fluctuations can be expected to be of order $\Omega^{1/2}$ and reflect the demographic stochasticity.

At any given time, the state of the full system is completely described by the state $\sigma$ of the environment and the number of individuals in the population $n$. 
We restrict ourselves to cases in which the environment has two states, $\sigma\in\{0,1\}$. The switching between these states is assumed to be independent of the state of the population $n$ and it occurs with constant rates. We write $\lambda_+$ for the rate of switching from state $0$ to state $1$, and $\lambda_-$ for the rate of switching from state $1$ to state $0$.

This stylised model can be interpreted in the context of genetic networks as follows \cite{kepler2001stochasticity}. The two states of the environment, often labelled as $\mathcal{G}_0$ and $\mathcal{G}_1$, correspond to regimes in which a certain promoter---a region on DNA which initiates transcription---is either inactive ($\mathcal{G}_0$) or activated ($\mathcal{G}_1$). A protein ($\mathcal{A}$) is produced with rates $b_0$ and $b_1$ in the corresponding state. Independently of the state of the gene, proteins degrade at a constant rate $d$. This simple model has been studied for example in \cite{kepler2001stochasticity, grima2012steady, thomas2014phenotypic, duncan2015noise, zeiser2010autocatalytic}, and it can be written in the form
\begin{align*} 
\mathcal{G}_0 &\xrightarrow{\mathmakebox[0.5cm]{\Omega b_0}} \mathcal{G}_0 + \mathcal{A}, \\
\mathcal{G}_1 &\xrightarrow{\mathmakebox[0.5cm]{\Omega b_1}} \mathcal{G}_1 + \mathcal{A}, \\
\mathcal{G}_0 &\xrightarrow{\mathmakebox[0.5cm]{\lambda_+}} \mathcal{G}_1, \nonumber \\
\mathcal{G}_1 &\xrightarrow{\mathmakebox[0.5cm]{\lambda_-}} \mathcal{G}_0, \nonumber\\
\mathcal{A} &\xrightarrow{\mathmakebox[0.5cm]{d}} \varnothing.
\numberthis
\label{eq:simple_reactions_all}
\end{align*}
The reaction rates of this model are linear in $n$. It is possible to develop exact solutions to linear models of this type using a generating-function approach \cite{peccoud1995markovian, hornos2005self, raj2006stochastic, shahrezaei2008analytical, grima2012steady, kumar2014exact}. However, such solutions are often limited to simple model systems and frequently they only provide limited insight into the actual physical dynamics. We use this linear model to introduce our approximation method. In later sections we will then apply this approach to cases with nonlinear reaction rates, where an exact solution is no longer feasible.

\begin{figure*}
\captionsetup[subfigure]{labelformat=empty}
{\centering
\subfloat{\includegraphics[height=0.42\columnwidth]{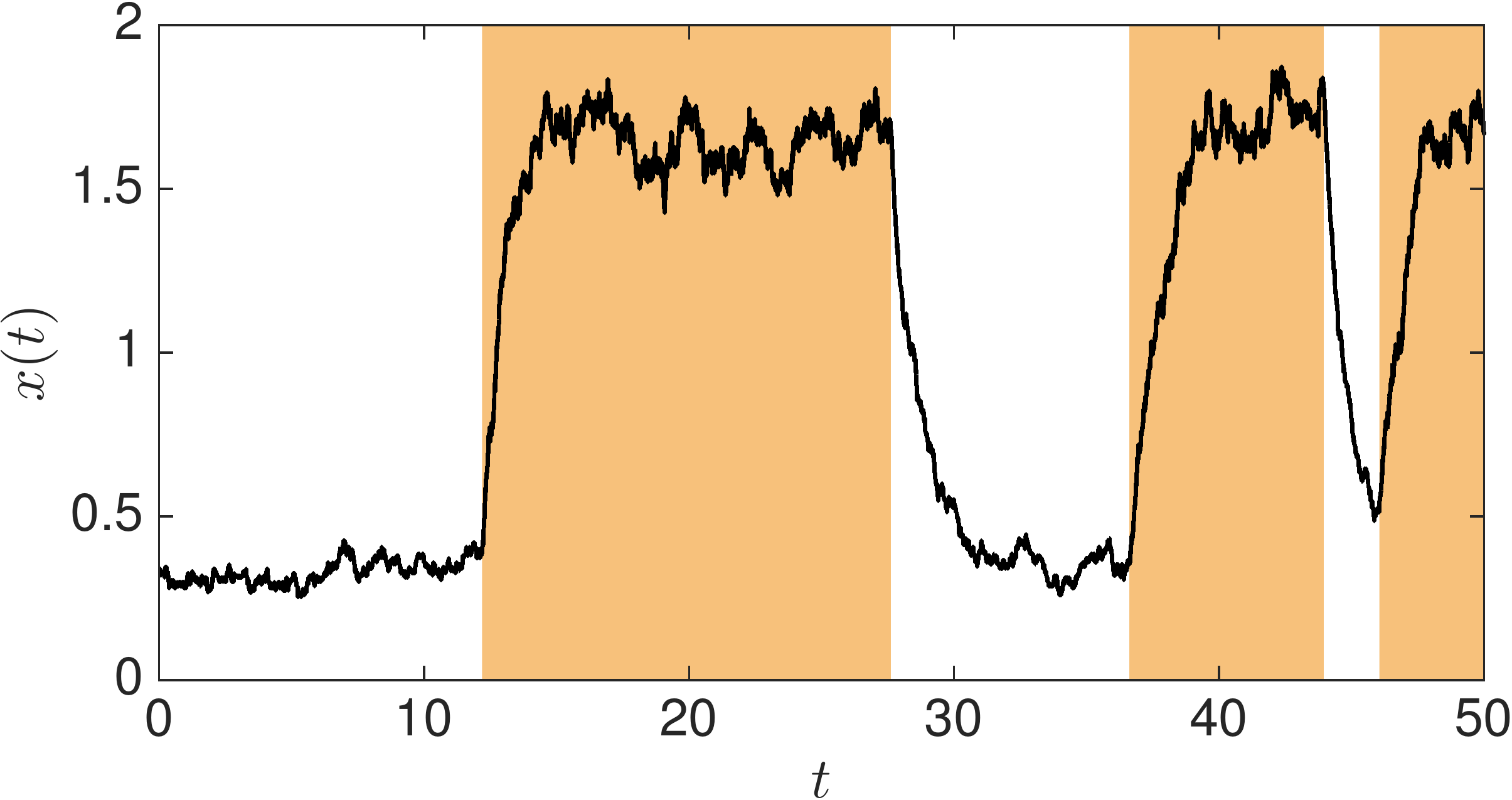}}\quad\quad ~
\subfloat{\includegraphics[height=0.42\columnwidth]{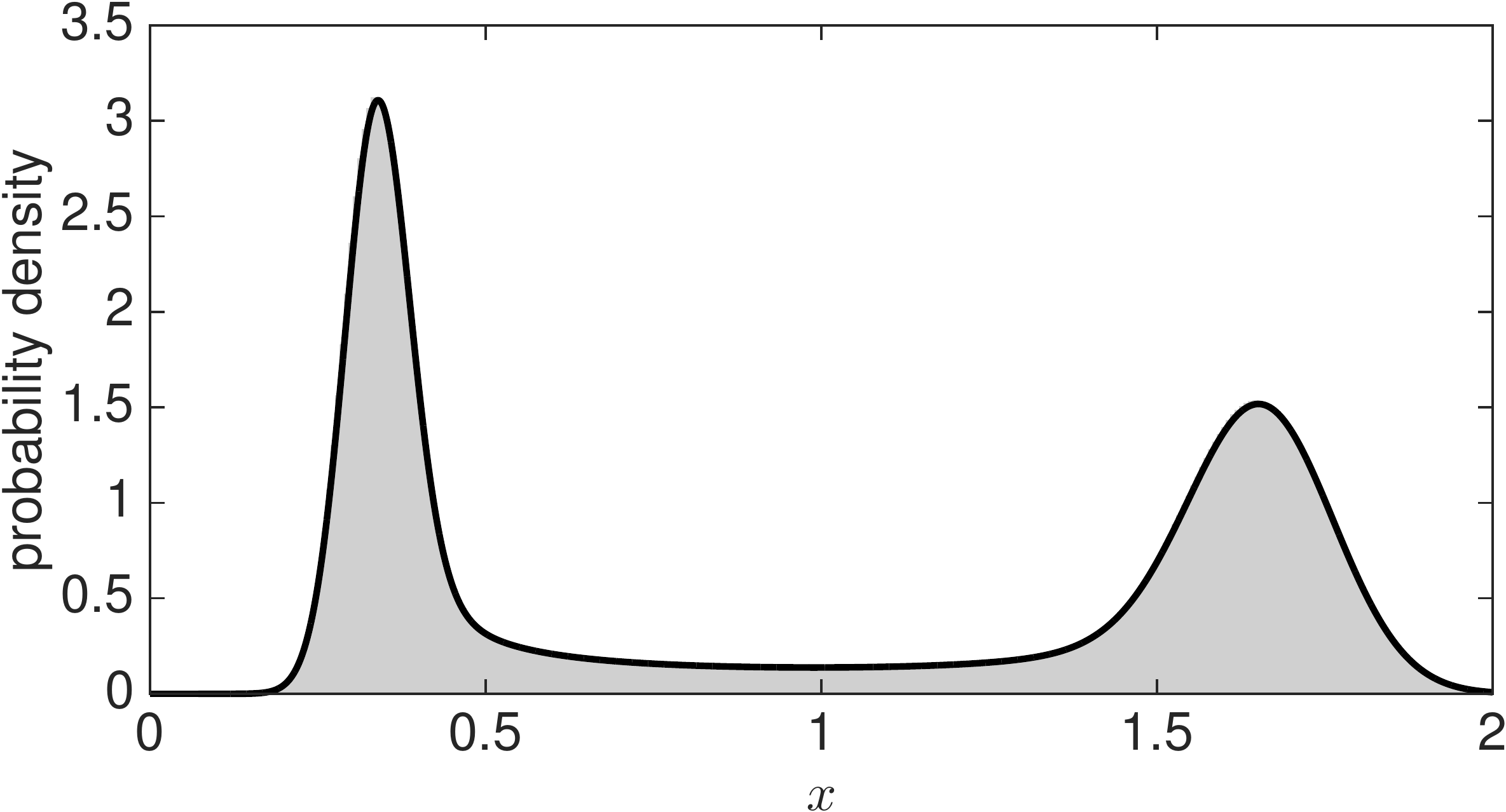}} \\
{\footnotesize (a) Slow switching ($\lambda_{+}=\lambda_{-}=0.1$).}\\
\subfloat{\includegraphics[height=0.42\columnwidth]{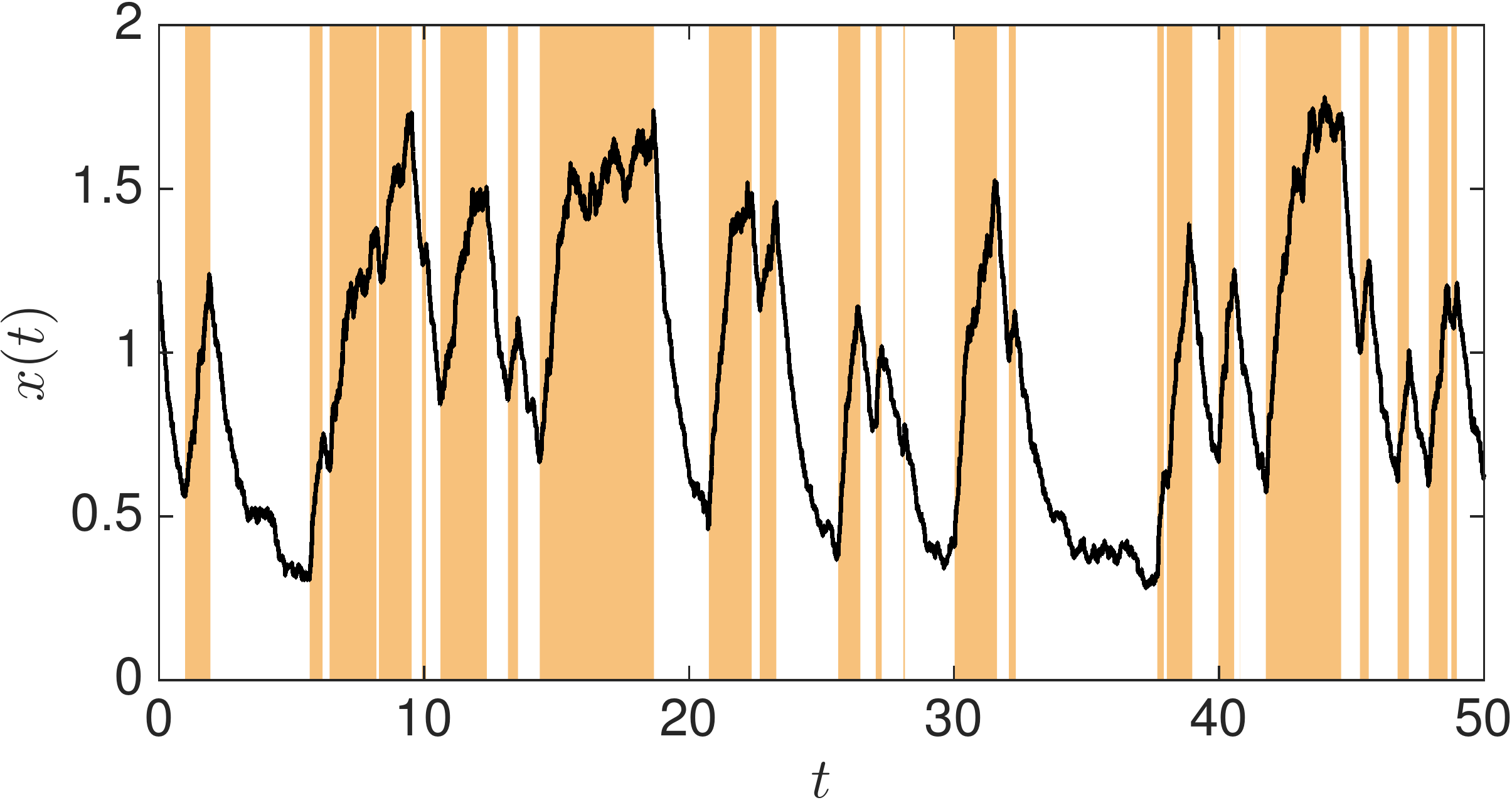}}\quad\quad ~
\subfloat{\includegraphics[height=0.42\columnwidth]{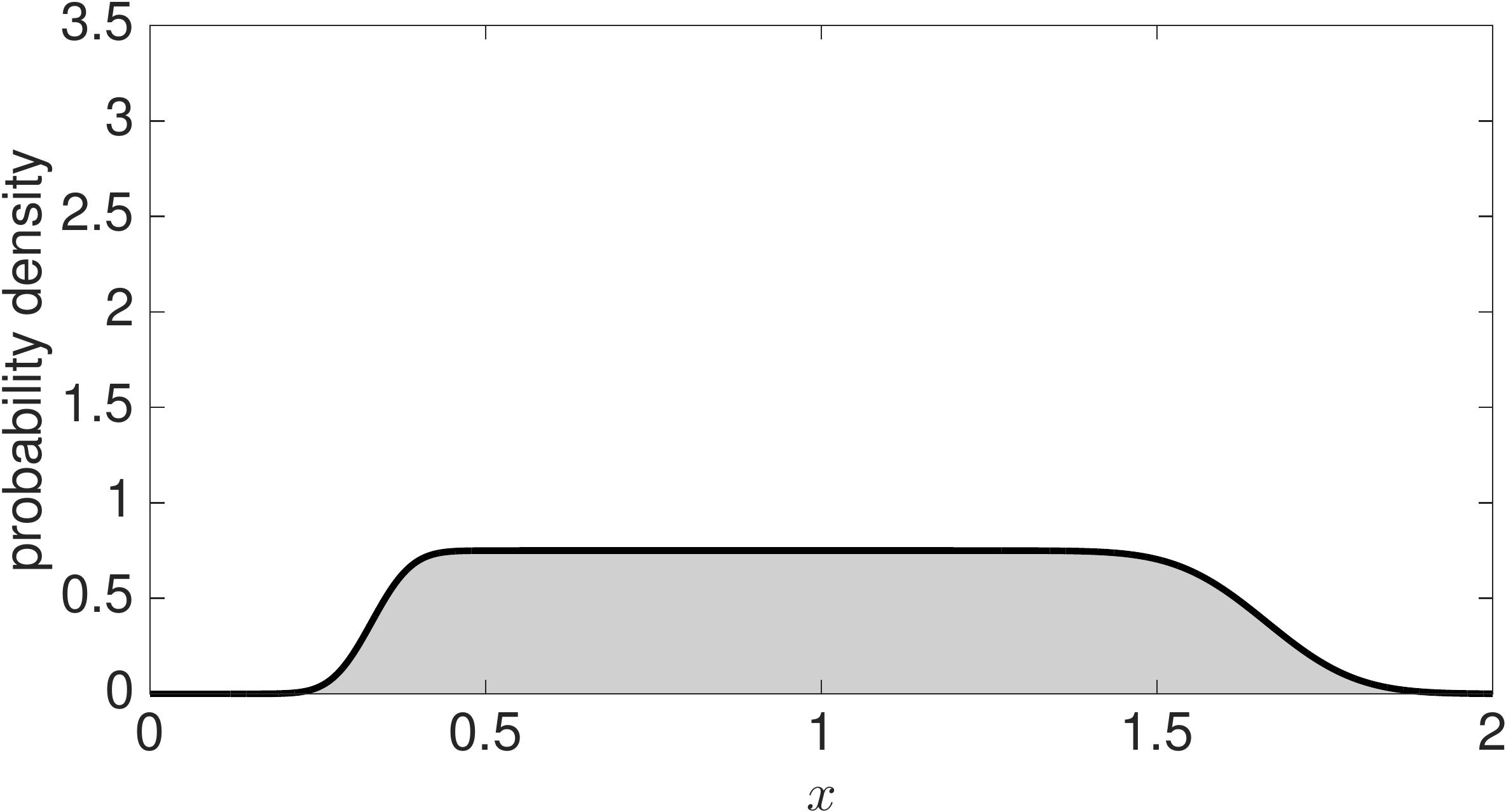}} \\
{\footnotesize (b) Intermediate rates of switching ($\lambda_{+}=\lambda_{-}=1$).} \\
\subfloat{\includegraphics[height=0.42\columnwidth]{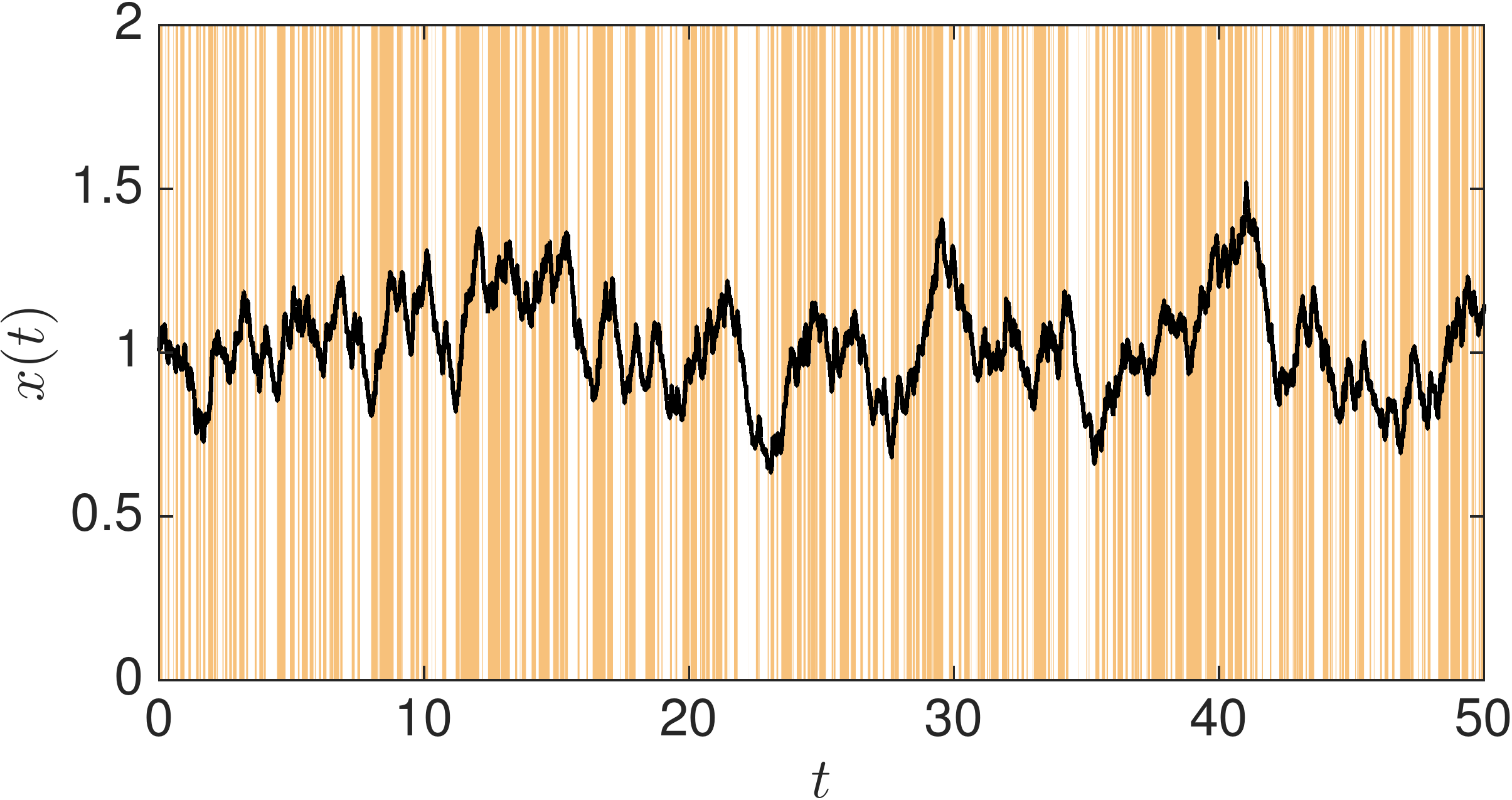}}\quad\quad ~
\subfloat{\includegraphics[height=0.42\columnwidth]{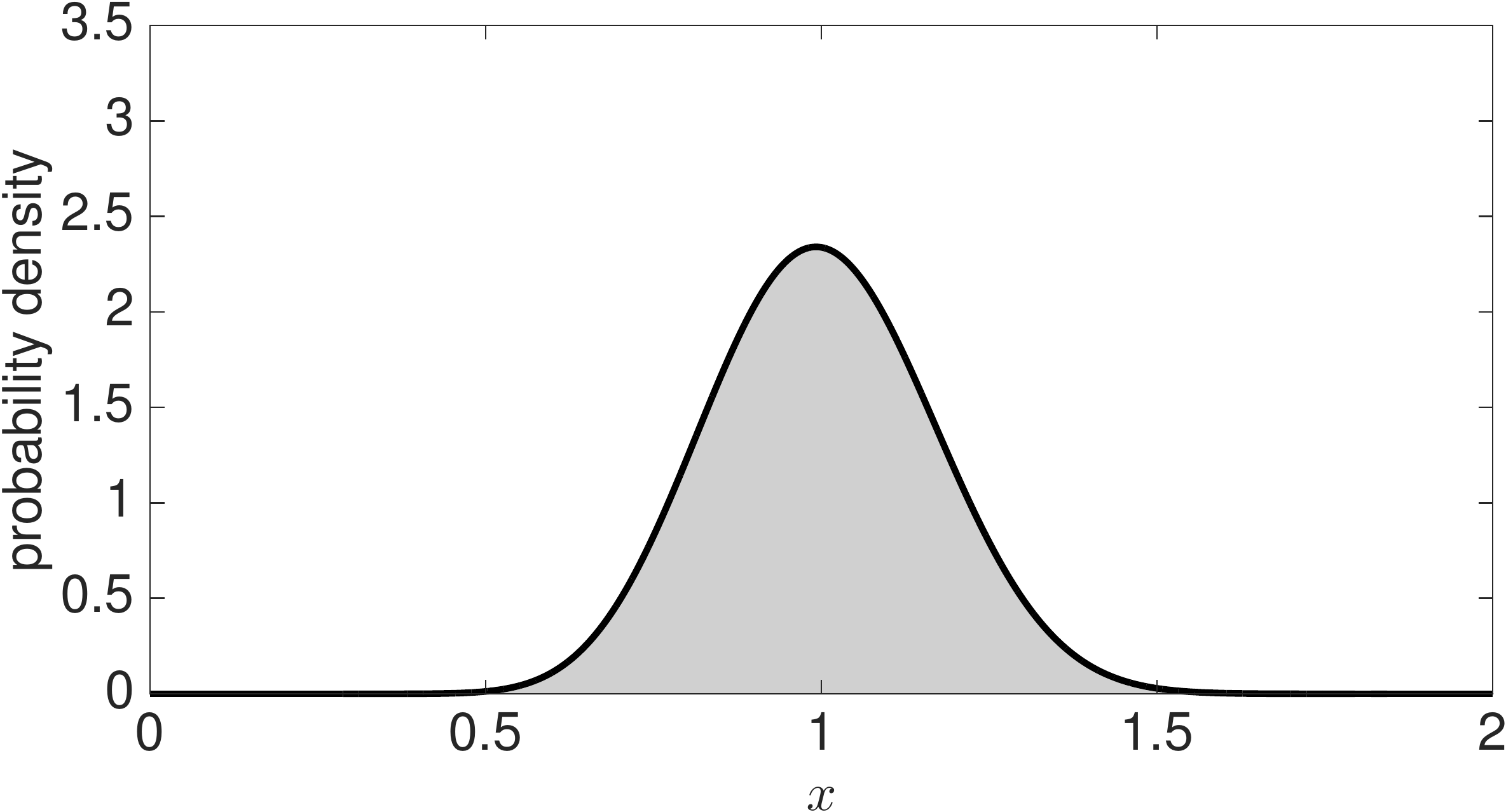}} \\
{\footnotesize (c) Fast switching ($\lambda_{+}=\lambda_{-}=10$).} 
}
\caption[Many]{Sample paths of the dynamics and the stationary distribution of the linear birth-death process described by Eq.~(\ref{eq:simple_reactions_all}). The left-hand panels show individual trajectories (solid line) and the state of the environment (shaded background). The panels on the right show the stationary distribution from simulations, averaged over multiple runs (histogram). We also plot the theoretical prediction for the stationary distribution $\Pi^*(x)$ obtained from Eqs.~\eqref{eq:full}, shown as solid lines for the three scenarios. The trajectories and stationary distributions have been obtained by application of the Gillespie algorithm \cite{gillespie1976general, gillespie1977exact}. Model parameters are $b_0=1/3$, $b_1=5/3$, $d=1$ and $\Omega=150$.}
\label{fig:simple_figs_x}
\end{figure*}

\subsection{Simulation of results and general behaviour}
To illustrate the general behaviour of the model, we first show the outcome of a set of characteristic simulations in Fig.~\ref{fig:simple_figs_x}. The left-hand panels depict individual simulation runs. In each panel on the left a trajectory of the population density $x(t)=n(t)/\Omega$ is shown as a solid line, and the switching of the environmental states is indicated by the shading of the background.
The upper panel shows an example of relatively slow switching. In each environmental state, the population tends to a fixed point, $\phi^*_\sigma=b_\sigma/d$, specific to the environment. It then fluctuates about this fixed point. The stationary distribution (histogram on the right) is bimodal. As the switching rate is increased (middle row), the stochastic dynamics spends more time in between the two fixed points and the bimodality of the stationary distribution is lost; we observe a nearly flat distribution between the two fixed points. At very fast switching (lower row) the stationary distribution becomes unimodal, peaked at a value between the two fixed points. The system spends most of its time fluctuating about a point in the interior of phase space, away from either of the two fixed points.
 
We set out to characterise this behaviour analytically, and our aim is to approximate the stationary outcome of the dynamics. We will develop a general approximation technique, which is applicable to a wider class of models, in the next section. Before we do this, it is useful to outline our approach and to describe the main steps of the analysis in the simpler model defined in Eqs.~\eqref{eq:simple_reactions_all}.

\subsection{Master equation}
We write $n(t)$ for the number of individuals at time $t$, and $\sigma(t)$ for the state of the environment. These follow random-jump Markov processes \cite{gardiner1985handbook}. We write $P(n,\sigma, t)$ for the probability to find the system and environment in state $(n,\sigma)$ at time $t$. We will frequently suppress the explicit time dependence to keep the notation compact.

The master equation governing the time evolution of this distribution can then be written as
\begin{align*}
\frac{\d}{\d t} P(n,0)={}&L_0 P(n,0)- \lambda_{+} P(n,0)+ \lambda_{-} P(n,1), \\
\frac{\d}{\d t} P(n,1)={}&L_1 P(n,1)+ \lambda_{+} P(n,0) - \lambda_{-} P(n,1)
.
\numberthis
\label{eq:simple_master_equation_n}
\end{align*}
These equations consist of two components. The operators $L_0$ and $L_1$ characterise the creation and removal of individuals assuming a fixed state of the environment. They are given by
\begin{align*}
L_\sigma= \Omega b_\sigma \left( E^{-1}-1 \right) + d \left( E-1 \right) n,
\numberthis
\label{eq:simple_operators}
\end{align*}
where $E$ is the shift operator \cite{van1992stochastic}: $Ef(n)=f(n+1)$. It is important to note that operators, such as $E$ or $L_\sigma$, act on everything that follows to their right throughout our paper, e.g., $E nf(n)=(n+1)f(n+1)$. The latter two terms in the master equation describe the switching between the two environmental states. 

The master equation \eqref{eq:simple_master_equation_n} describes the time evolution of the full process, and in our analysis, we wish to calculate the joint stationary distribution of the number of individuals $n$ and environmental state $\sigma$, $P^*(n,\sigma)$.

\subsection{Approximation of the master equation}
We now proceed to approximate the above master equation. To this end, it is useful to define the population density $x(t)=n(t)/\Omega$. Assuming that the typical system size $\Omega$ is large but finite, the operators, $L_0$ and $L_1$, can be approximated by a Taylor expansion with respect to $\Omega^{-1}$,
\begin{align*}
L_\sigma\approx\mathcal{L}_\sigma=
- \partial_x (b_\sigma - xd) + \frac{1}{2\Omega} \partial_x^2 (b_\sigma+xd).
\numberthis
\label{eq:simple_operators_x}
\end{align*}
This is the Kramers--Moyal expansion \cite{van1992stochastic, gardiner1985handbook}
in terms of the variable $x=n/\Omega$, while keeping the state of the environment, $\sigma$, discrete. We write $\mathcal{L}_\sigma$ for the operators obtained from this expansion, retaining only leading and sub-leading orders in $\Omega^{-1}$. The state of the system is now expressed in terms of $x$ and $\sigma$, and we will write $\Pi(x,\sigma)$ for the resulting probability density. The explicit time-dependence is again suppressed in this notation. 

Substituting this into the master equation \eqref{eq:simple_master_equation_n} allows us to approximate the process by
\begin{align*}
\partial_t \Pi(x,0)={}&\mathcal{L}_0 \Pi(x,0) - \lambda_{+} \Pi(x,0) + \lambda_{-} \Pi(x,1), \\
\partial_t \Pi(x,1)={}&\mathcal{L}_1 \Pi(x,1) + \lambda_{+} \Pi(x,0) - \lambda_{-} \Pi(x,1).
\numberthis
\label{eq:simple_master_equation_x}
\end{align*}
This expansion differs from the standard Kramers--Moyal expansion in that the environmental states are not included in the expansion. Equation~(\ref{eq:simple_master_equation_x}) accordingly is not a standard Fokker--Planck equation; it retains the discrete switching terms, akin to terms in the master equation of a conventional telegraph process. Equations~\eqref{eq:simple_master_equation_x} describe a diffusion process with a Markovian switching in between two sets of drift and diffusion \cite{mao2006stochastic, potoyan2015dichotomous}. 

Since Eqs.~\eqref{eq:simple_master_equation_x} contain multiplicative noise, it is difficult to solve these equations directly. In the following sections, we propose an approximation method to analyse this dynamics. Our approach is similar to the conventional linear-noise approximation, and describes the effects of intrinsic noise to sub-leading order.

\subsection{Leading-order approximation: piecewise-deterministic Markov process}
In the above expansion we have retained leading and sub-leading terms in $\Omega^{-1}$. To proceed, it is useful to first consider the leading-order terms only, i.e., to study the limit $\Omega\to\infty$. We obtain 
\begin{align*}
\partial_t \Pi(\phi,0)={}& - \partial_\phi (b_0 - d\phi)\Pi(\phi,0) \\
& - \lambda_{+} \Pi(\phi,0) + \lambda_{-} \Pi(\phi,1), \\
\partial_t \Pi(\phi,1)={}& - \partial_\phi (b_1 - d\phi)\Pi(\phi,1) \\
& + \lambda_{+} \Pi(\phi,0) - \lambda_{-} \Pi(\phi,1),
\numberthis
\label{eq:simple_pdmp_master}
\end{align*}
where we have written $\phi=\lim_{\Omega\to\infty} n/\Omega$ to indicate that we have taken the limit of infinite populations. In this limit the $\mathcal{L}_\sigma$ are Liouville operators describing deterministic flow of $\phi$. The functional form of this flow at any one time is entirely determined by the state of the environment, $\sigma$. The process $\phi(t)$ is a piecewise deterministic Markov process \cite{davis1984piecewise,faggionato2009non,bena2006dichotomous,mao2006stochastic}, a random process composed of deterministic motion in between the discrete environmental transitions. The PDMP is a description of the system which accounts for the stochasticity of the switching only; the demographic noise on the other hand is neglected in the above truncation after the leading-order term. 

\begin{figure}
\includegraphics[width=0.47\textwidth]{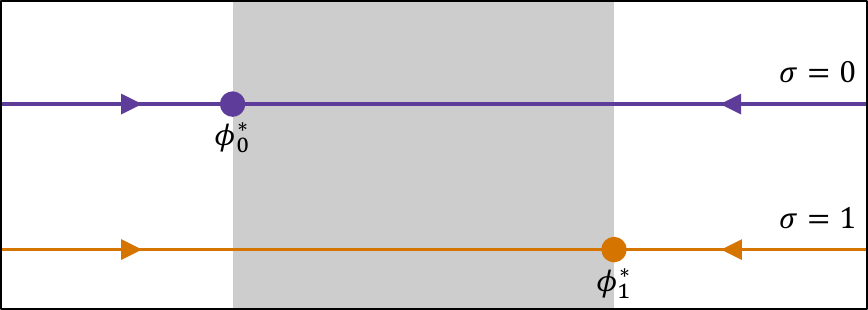}
\caption{Illustration of the Liouville flow of the model described by Eq.~(\ref{eq:simple_pdmp_master}). Arrows indicate the Liouville flow in each of the two environments, stable fixed points are shown as filled circles. The PDMP converges into the interval between the two fixed points (shaded region) at long times.}
\label{fig:simple_fps}
\end{figure}
The long-term behaviour of the system can be determined from inspection of the Liouville operators. In each state $\sigma$ the process tends towards a stable fixed point, $\phi_\sigma^*=b_\sigma/d$. In the following, we assume that $b_0<b_1$ so that $\phi_0^*<\phi_1^*$. The dynamics can be illustrated by the flow diagram in Fig.~\ref{fig:simple_fps}. When the environment is in state $0$ the trajectory of the PDMP moves towards $\phi_0^*$, and when the environment switches the direction is reversed towards $\phi_1^*$. In the long-run, the PDMP will always take values in the interval between the fixed points $\phi_0^*$ and $\phi_1^*$.

\begin{figure*}
\captionsetup[subfigure]{labelformat=empty}
{\centering
\subfloat{\includegraphics[height=0.42\columnwidth]{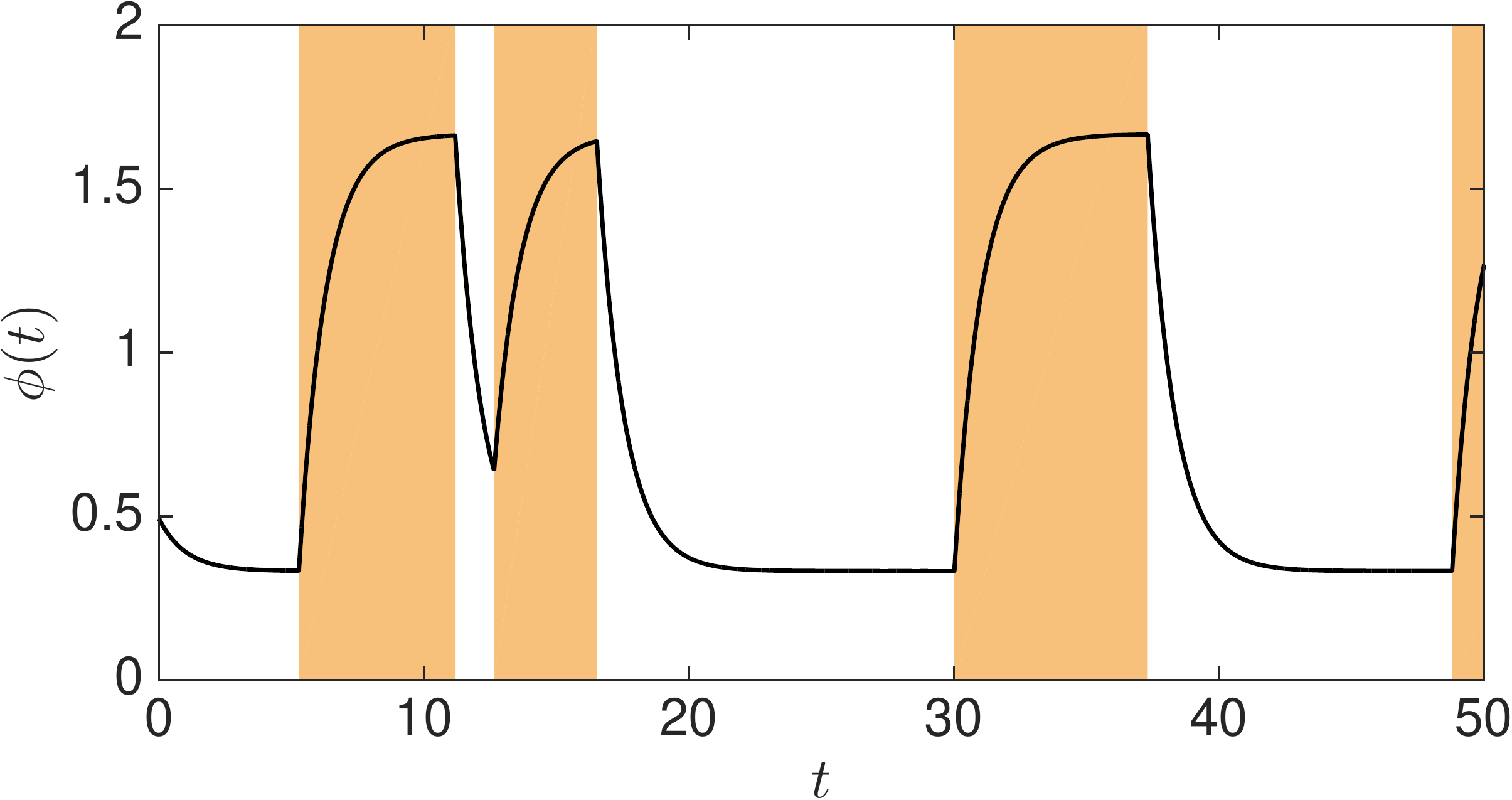}}\quad\quad ~
\subfloat{\includegraphics[height=0.42\columnwidth]{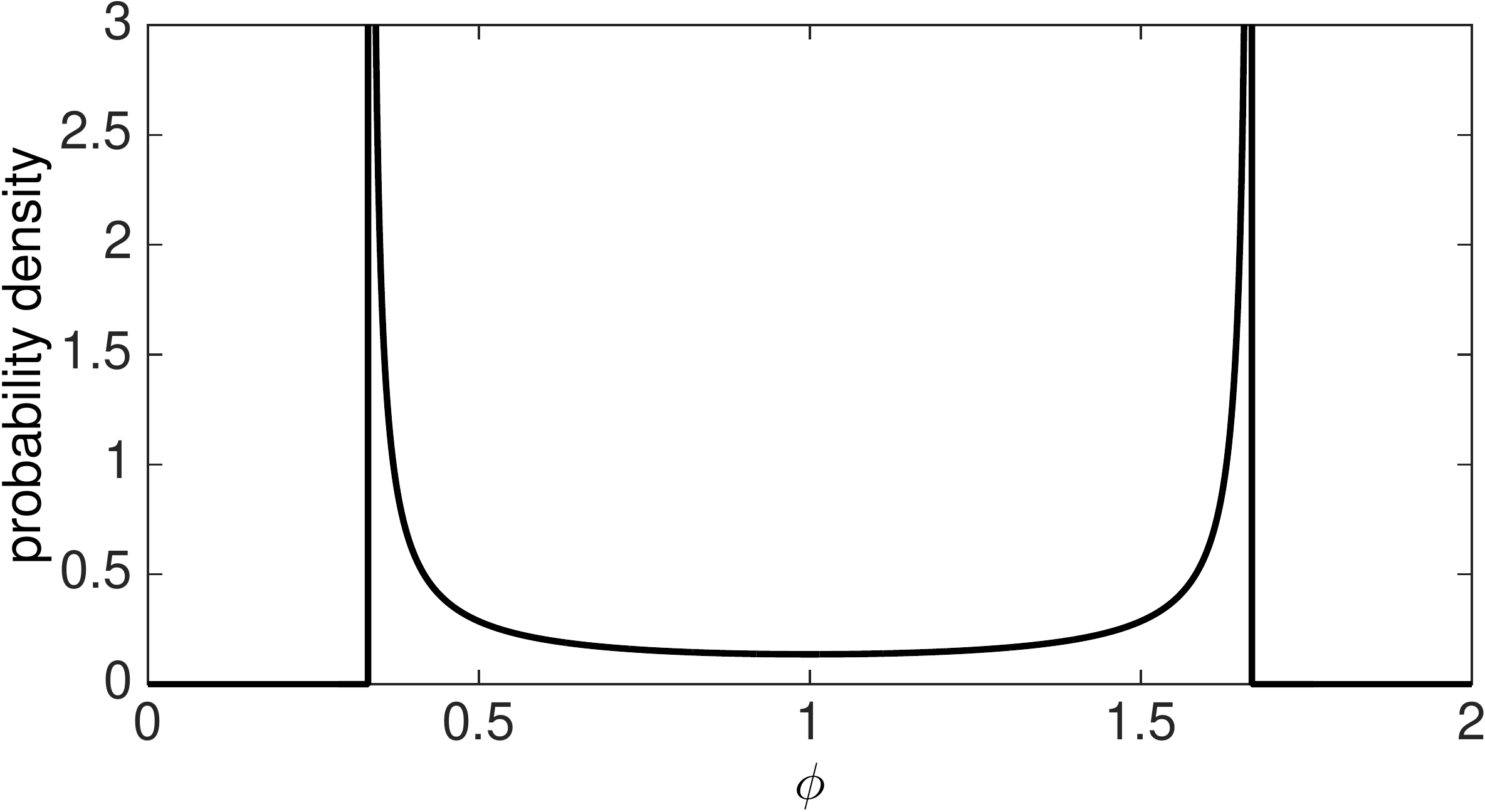}} \\
{\footnotesize (a) Slow switching ($\lambda_{+}=\lambda_{-}=0.1$).}\\
\subfloat{\includegraphics[height=0.42\columnwidth]{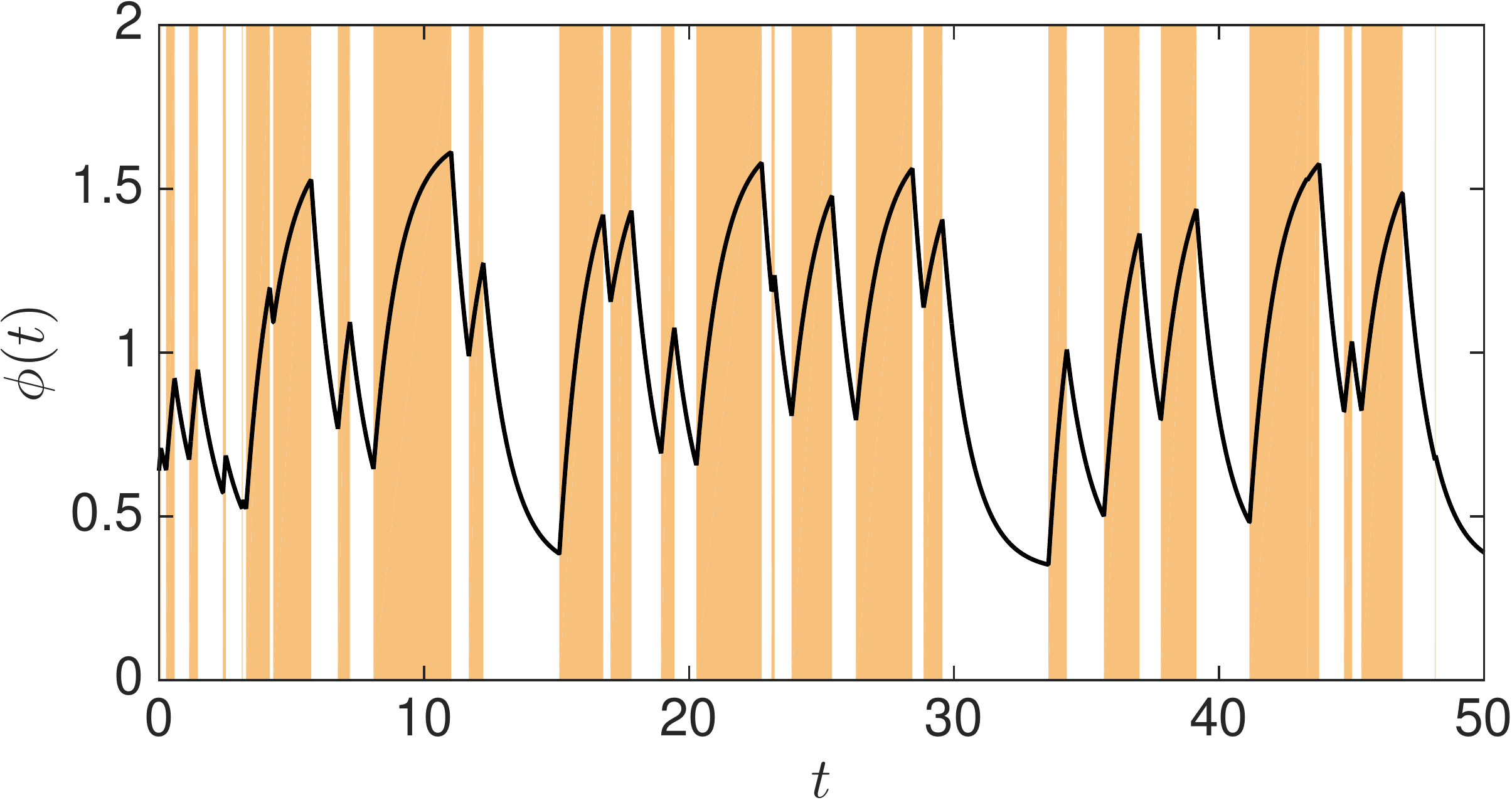}}\quad\quad ~
\subfloat{\includegraphics[height=0.42\columnwidth]{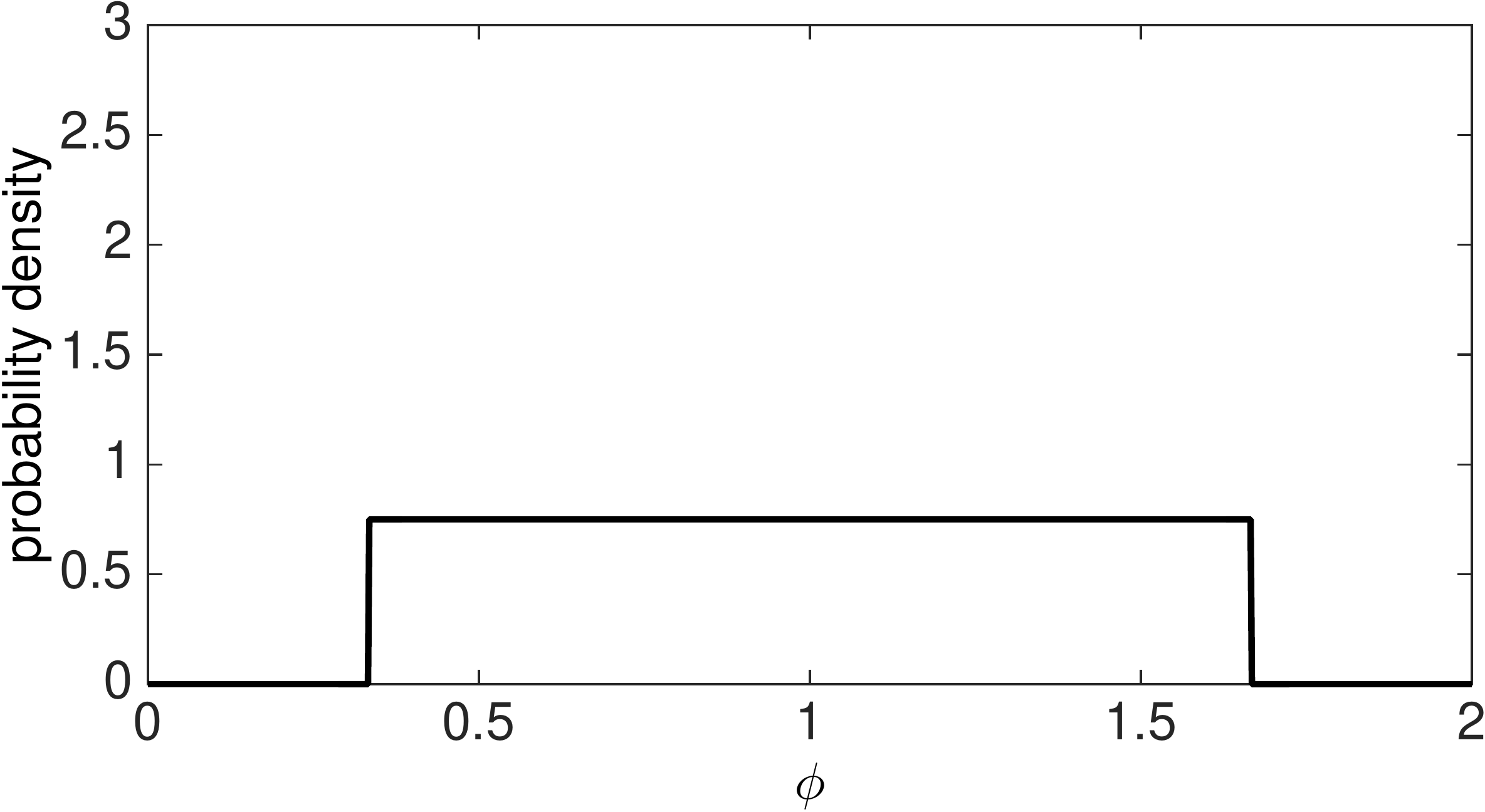}} \\
{\footnotesize (b) Intermediate switching rates ($\lambda_{+}=\lambda_{-}=1$).} \\
\subfloat{\includegraphics[height=0.42\columnwidth]{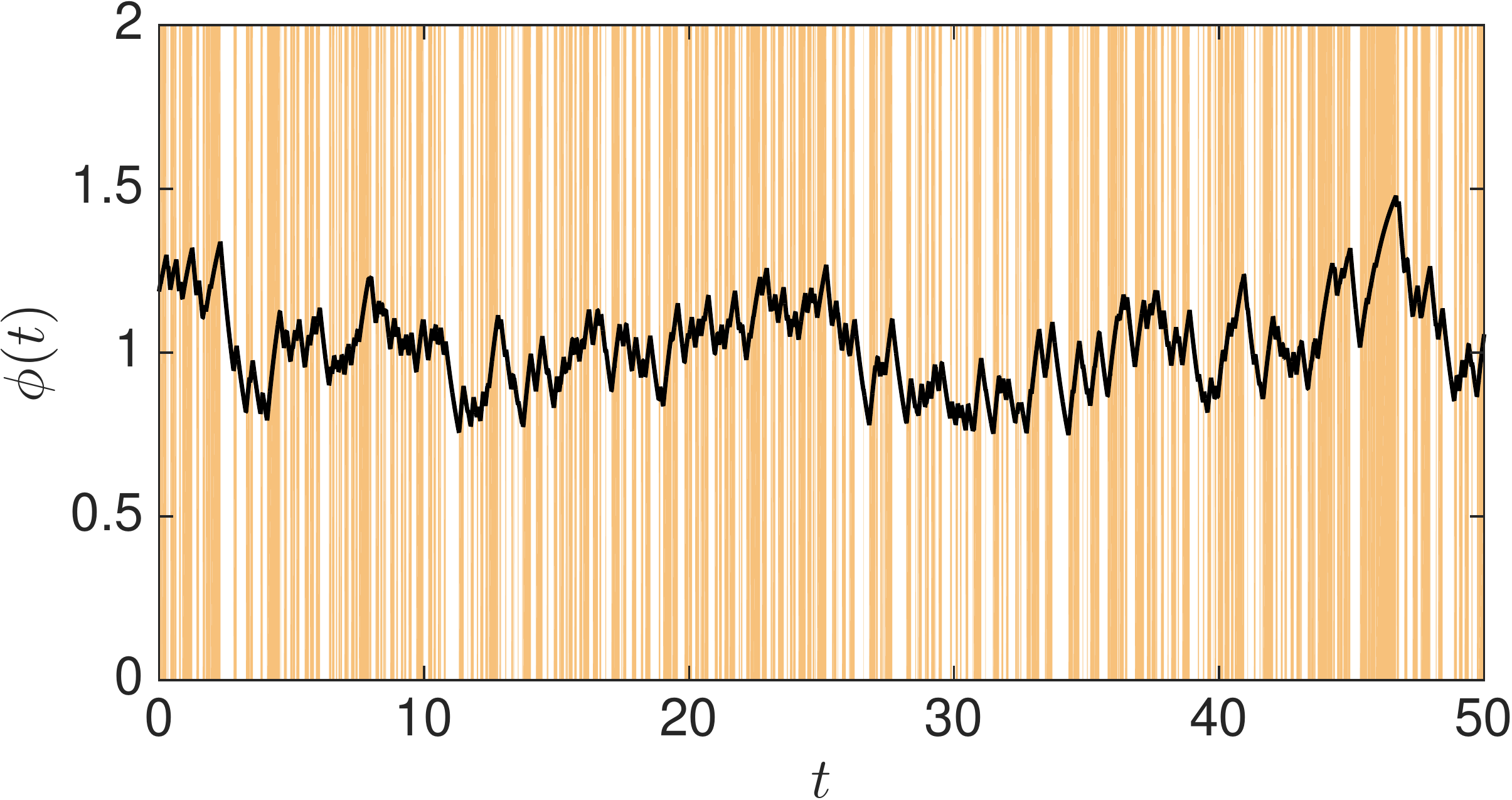}}\quad\quad ~
\subfloat{\includegraphics[height=0.42\columnwidth]{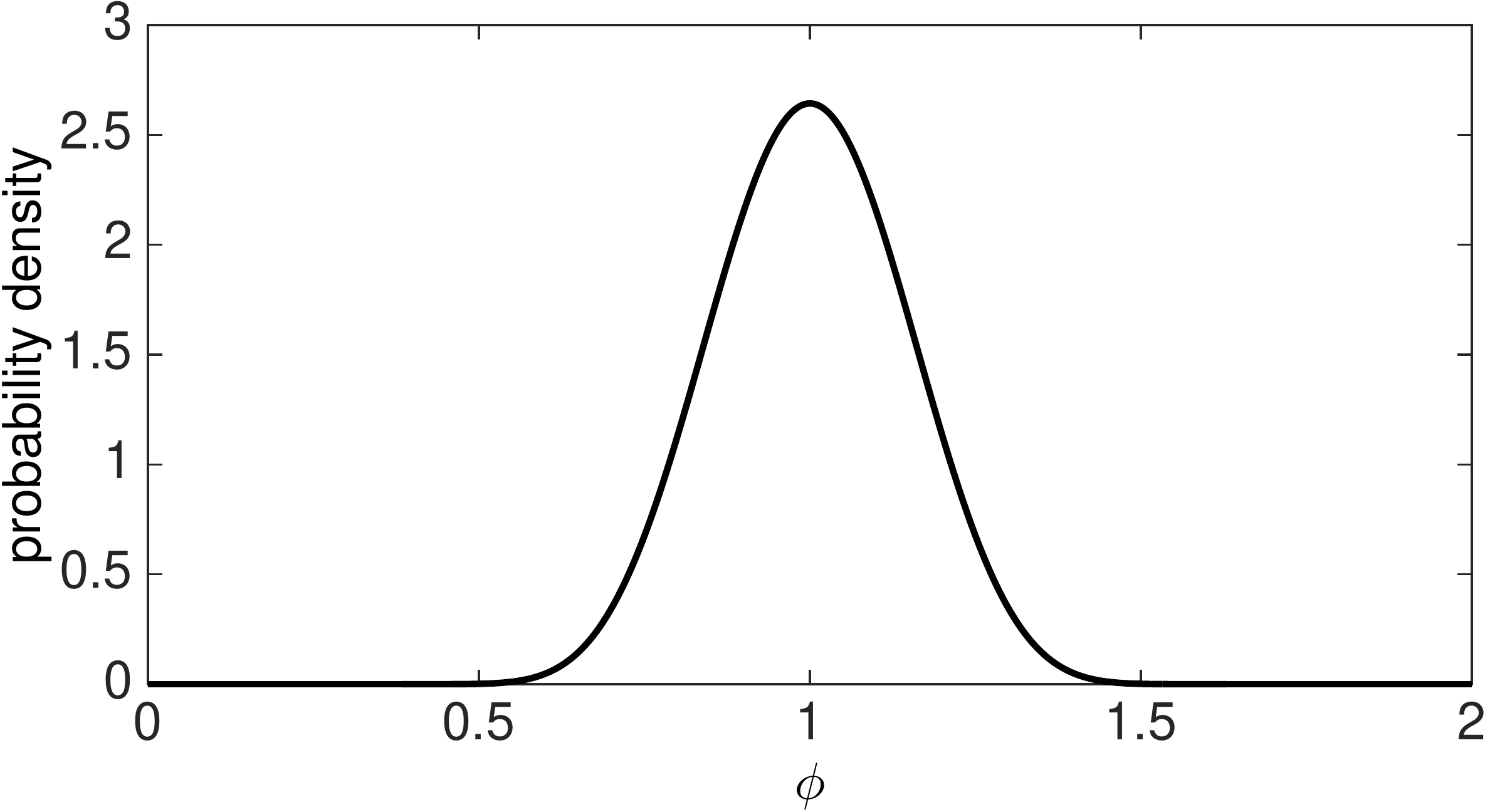}} \\
{\footnotesize (c) Fast switching ($\lambda_{+}=\lambda_{-}=10$).} 
}
\caption[Many]{Sample paths and stationary distribution of the PDMP for the linear model in infinite populations. Panels on the left show individual trajectories (solid line) and the state of the environment (shaded background). The panels on the right show the stationary distribution of the PDMP, as obtained from from Eq.~\eqref{eq:simple_dist_phi}. Comparison with Fig.~\ref{fig:simple_figs_x} shows that this stationary distribution does not adequately reflect the stationary distribution of the full noisy process in finite populations. The sample paths have been obtained by Runge--Kutta integration of Eq.~\eqref{eq:pdmp0} between randomly-generated exponential switching times. Model parameters are $b_0=1/3$, $b_1=5/3$, $d=1$ and $\Omega=150$.}
\label{fig:simple_figs_phi}
\end{figure*}
The stationary distribution of the PDMP, $\Pi^*(\phi,\sigma)$, can be found by setting the derivatives with respect to time in Eqs.~\eqref{eq:simple_pdmp_master} to zero, followed by integration and rearrangement. Further details are discussed in a more general setting in the next section. For the linear model defined in Eqs.~\eqref{eq:simple_reactions_system} we find
\begin{align*}
\Pi^*(\phi,0)
&=\mathcal{N}\frac{
\left(\phi - \phi^*_0 \right)^{\frac{\lambda_{+}}{d}} 
\left(\phi^*_1 - \phi \right)^{\frac{\lambda_{-}}{d}}}{\phi-\phi^*_0},\\
\Pi^*(\phi,1)
&=\mathcal{N}\frac{
\left(\phi - \phi^*_0 \right)^{\frac{\lambda_{+}}{d}} 
\left(\phi^*_1 -\phi\right)^{\frac{\lambda_{-}}{d}}}{\phi^*_1-\phi}
,
\numberthis
\label{eq:simple_dist_phi}
\end{align*}
for $\phi \in (\phi_0^*, \phi_1^*)$. The pre-factor $\mathcal{N}$ is a normalisation constant, determined by the condition
\begin{align*}
\int_{\phi^*_0}^{\phi_1^*} \left[\Pi^*(\phi,0)+\Pi^*(\phi,1)\right] \, \d \phi =1.
\numberthis
\label{eq:simple_normalisation}
\end{align*}
This result is consistent with those reported by other authors \cite{zeiser2010autocatalytic,bena2006dichotomous,kitahara1979coloured}.

\subsection{Comparison against simulations}
In Fig.~\ref{fig:simple_figs_phi} we illustrate the behaviour of the PDMP. The panels on the left show individual time series for slow, medium and fast switching of the environment (parameters are as in Fig.~\ref{fig:simple_figs_x}). The panels on the right depict the corresponding stationary distributions of the PDMP, as obtained from Eqs.~\eqref{eq:simple_dist_phi}. For slow switching rates, the marginal stationary distribution $\Pi^*(\phi)=\Pi^*(\phi,0)+\Pi^*(\phi,1)$ is bimodal, with singularities at the endpoints. In this regime, the PDMP typically spends enough time in each environment between switches to come close to the corresponding fixed point. For fast switching $\Pi^*(\phi)$ is unimodal. In this situation, the system typically does not have sufficient time to reach the vicinity of the fixed points. An intermediate case is shown in the middle panels. For this particular choice of parameters, the resulting stationary distribution is seen to be flat between the two fixed points, $\phi_0^*$ and $\phi_1^*$.

Comparison of the stationary distributions of the PDMP with those of the process in finite populations (Fig.~\ref{fig:simple_figs_x}) shows that the PDMP approximation recovers some of the qualitative features of the full system, but not all. The transition from a bimodal to a unimodal shape is successfully reproduced. On the other hand, the singularities for slow switching rates seen in the stationary state of the PDMP are not observed in the full process. The PDMP is confined to the interval $(\phi^*_0,\phi^*_1)$, while the intrinsic noise in finite populations allows for fluctuations on both sides of $\phi_0^*$ and $\phi_1^*$. Thus the support of the stationary distribution of the model with intrinsic noise includes concentrations below $\phi_0^*$ and above $\phi_1^*$.

These results stress the significance of intrinsic noise for the dynamics of the system. In order to approximate the stationary behaviour to a better accuracy, it is necessary to include higher-order terms in the above Kramers--Moyal expansion. The corresponding formalism is well established for systems without random switches of the environment, and it takes the form of an expansion about the deterministic path of the infinite system \cite{boland2009limit}. In our case, the leading-order behaviour (the PDMP) is a stochastic process itself due to the randomness of the environmental switching. The sub-leading description we discuss below is hence an expansion about this random process.

\subsection{Sub-leading order: linear-noise approximation}
Before we present a more detailed account of the expansion to sub-leading order (Section~\ref{sec:general}), we briefly outline the general idea. For a fixed realisation of the environmental switching process, $\sigma(t)$, we decompose the dynamics of the population as follows:
\begin{align}
\frac{n}{\Omega}=\phi(t)+\frac{1}{\sqrt{\Omega}}\xi(t).
\end{align}
This is along the lines of the system-size expansion in systems with time-dependent rates \cite{boland2008limit,boland2009limit}. For a given path of the environment, $\sigma(t)$, the trajectory $\phi(t)$ of the resulting PDMP is given by the solution of
\begin{align}
\dot \phi(t)=b_{\sigma(t)}-d\phi(t),
\label{eq:pdmp0}
\end{align}
where the birth rate $b_{\sigma(t)}$ at time $t$ is determined by the state of the environment at that time, $\sigma(t)$. 
The dynamics of Eq.~\eqref{eq:pdmp0} is the equivalent of the usual rate equations for systems without environmental switching.

To sub-leading order, systems of this form, but without environmental switching, are described by stochastic differential equations of the form
\begin{align}
\dot x(t) = v(x)+\sqrt{\frac{w(x)}{\Omega}}\eta(t),
\label{eq:sde}
\end{align}
where we have suppressed the obvious time dependence of $x$ on the right-hand side. The dependence of the amplitude $w(x)$ on $x$ indicates multiplicative noise, and $\eta(t)$ is Gaussian white noise of unit amplitude, i.e., $\avg{\eta(t)\eta(t')}=\delta(t-t')$. The term $v(x)$ represents deterministic drift. In a birth-death process with fixed birth rate $b$ and death rate $d$ one would have $v(x)=b-dx$, for example.

The equivalent of Eq.~\eqref{eq:sde} for the model with environmental switching is given by
\begin{align}
\dot x(t)=v_{\sigma(t)}(x)+\sqrt{\frac{w_{\sigma(t)}(x)}{\Omega}}\eta(t).
\label{eq:psde}
\end{align}
Eq.~\eqref{eq:psde} is similar to the outcome of the procedure described in \cite{boland2008limit,boland2009limit}, where the system-size expansion was carried out for systems with periodically driven rates. The main difference is the fact that $\sigma(t)$ is now a stochastic process itself, whereas the external driving was deterministic in \cite{boland2009limit}. In the limit $\Omega\to\infty$ the noise $\eta(t)$ does not contribute, and we recover the PDMP dynamics of Eq.~\eqref{eq:pdmp0}.

For the case of the linear model described above we have $v_\sigma(x)=b_\sigma-dx$ and $w_\sigma(x)=b_\sigma+dx$. These are the drift term and noise amplitude one would obtain from a standard Kramers--Moyal expansion at fixed environmental state $\sigma$.

In the spirit of the usual linear-noise approximation (LNA) we next write $x(t)=\phi(t)+\xi(t)/\sqrt{\Omega}$ and obtain
\begin{eqnarray}
\dot\phi(t) & = & v_{\sigma(t)}(\phi), \nonumber \\
\dot\xi(t) & = & v'_{\sigma(t)}(\phi)\xi(t)+\sqrt{w_{\sigma(t)}(\phi)}\eta(t).
\end{eqnarray}
We have written $v'_\sigma=\d v_\sigma(\phi) / \d \phi$, and we have suppressed the time dependence of $\phi$ on the right-hand side.

From the LNA it is possible to proceed and to approximate the stationary distribution of the process, $\Pi^*(\xi,\phi,\sigma)=\Pi^*(\xi|\phi,\sigma)\Pi^*(\phi,\sigma)$. The distribution $\Pi^*(\phi,\sigma)$ is the stationary outcome of the PDMP, and it can be computed exactly, see Eqs.~\eqref{eq:simple_dist_phi} for the linear birth-death model. The general case is discussed in the following section.

\begin{figure}
\includegraphics[width=1\columnwidth]{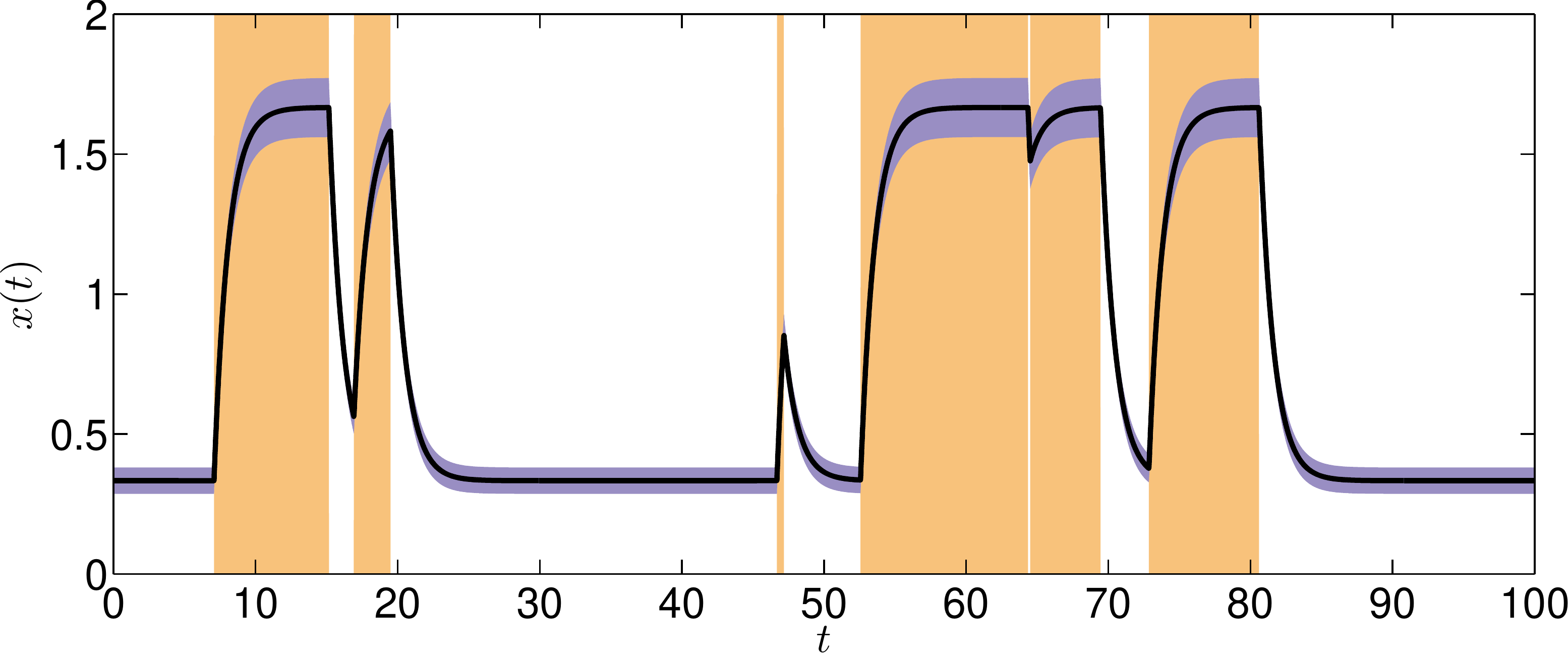}
\caption{Sample path of the PDMP of the linear model ($\phi(t)$, solid line) and environmental state $\sigma(t)$ (background shading). The dynamics in the finite system for the same realisation of the environmental process will deviate from the PDMP. The standard deviation of the deviation is approximated by Eq.~\eqref{eq:app_noise_variance} as discussed below, and shown here as shading around the PDMP trajectory. Parameters are $b_0=1/3$, $b_1=5/3$, $d=1$, $\lambda_{+}=\lambda_{-}=1$ and $\Omega=150$.}
\label{fig:simple_xi_sample}
\end{figure}

In the stationary regime the distribution of $\xi$ will, in principle, depend on the state $\sigma$ of the environment and on the variable $\phi$. In order to proceed we now assume that the dependence on $\sigma$ can be neglected, so that we write $\Pi^*(\xi|\phi,\sigma)\approx \Pi^*(\xi|\phi)$. This is an approximation, but it turns out to work well for all models we have tested. Making this assumption we write
\begin{align}
\Pi^*(\xi,\phi,\sigma)\approx \Pi^*(\xi|\phi)\ \Pi^*(\phi,\sigma).
\end{align}
The distribution $\Pi^*(\xi|\phi)$ is Gaussian with mean zero, and with a variance which depends on $\phi$ and which can be obtained analytically as described in the next section. The resulting picture is illustrated in Fig.~\ref{fig:simple_xi_sample}. A given realisation of the environmental process generates a realisation of the PDMP. The state of a finite population, subject to the same path of environmental states, will fluctuate about the PDMP trajectory, as indicated by the shading in Fig.~\ref{fig:simple_xi_sample}. 

From these approximations the stationary distribution of $x=\phi+\Omega^{-1/2}\xi$ can then be estimated as
\begin{align*}
\Pi^*(x)=\sum_\sigma \int \d\phi \,\d\xi~\bigg[{}& \Pi^*(\xi|\phi)\Pi^*(\phi,\sigma)\\
&\times \delta(x-\phi-\Omega^{-1/2}\xi)\bigg].
\numberthis
\label{eq:full}
\end{align*}

Returning to Fig.~\ref{fig:simple_figs_x}, we compare the stationary distribution of the linear model as obtained from Eq.~\eqref{eq:full} against the results of numerical simulations. The results from the theory are shown as solid lines, and the simulation data as histograms. We find that the approximation reproduces the numerical results to a good accuracy.

\section{General formalism}
\label{sec:general}
\subsection{Definition and master equation}
The linear model discussed so far was deliberately simple and the main purpose of studying it was to develop a general intuition. In order to extend the method beyond the linear case, we now consider a more general model. This will introduce several new aspects to the problem.

As before, we restrict our discussion to the case of a single species and two environmental states. We write $n$ for the number of individuals in the population, and $\sigma\in\{0,1\}$ for the state of the environment. We use the notation $\lambda_+(n)$ for the rate with which the environment switches from state $0$ to state $1$, and $\lambda_-(n)$ for the rate of switches in the opposite direction. Unlike in the previous sections, we now allow for an explicit dependence of these rates on the state of the population, $\lambda_\pm=\lambda_\pm(n)$.

We assume that there are $M$ possible reactions in the population, labelled $m=1,\dots,M$. Each reaction $m$ occurs with rate $a_{m,\sigma}(n)$ dependent on the current state of the environment and population. Any occurrence of a reaction of type $m$ is taken to change the number of individuals in the population by $S_m$. These are the underlying stoichiometric coefficients. The set of propensity functions, $a_{m,\sigma}(n)$, together with the stoichiometric coefficients completely define the dynamics of the population.

The master equation describing the time evolution of $P(n,\sigma)$ then reads 
\begin{align*}
\frac{\d}{\d t} P(n,0)={}&L_0 P(n,0) - \lambda_+(n) P(n,0) + \lambda_-(n) P(n,1), \\
\frac{\d}{\d t} P(n,1)={}&L_1 P(n,1) + \lambda_+(n) P(n,0) - \lambda_-(n) P(n,1).
\numberthis
\label{eq:general_master_equation}
\end{align*}
The operators $L_0$ and $L_1$ are given by
\begin{align*}
L_\sigma=& \sum_{m=1}^M \left( E^{-S_{m}}-1 \right) a_{m,\sigma}(n).
\numberthis
\label{eq:general_operators}
\end{align*}

Anticipating the system-size expansion, we write the transition rates in the form $a_{m,\sigma}(n)=\Omega r_{m,\sigma}(x)$, where $x=n/\Omega$ and where $r_{m,\sigma}$ carries no explicit $\Omega$-dependence. In slight abuse of notation we will write $\lambda_\pm(x)$ instead of $\lambda_\pm(\Omega x)$. 
The master equation can be expressed in terms of $\Pi(x,\sigma,t)$, the probability density of finding the random processes at $x,\sigma$ at time $t$. The dynamics of the joint probability distribution can be approximated by
\begin{align*}
\partial_t \Pi(x,0)=& \mathcal{L}_0 \Pi(x,0) 
- \lambda_+(x) \Pi(x,0) + \lambda_-(x) \Pi(x,1), \\
\partial_t \Pi(x,1)=& \mathcal{L}_1 \Pi(x,1) 
+ \lambda_+(x) \Pi(x,0) - \lambda_-(x) \Pi(x,1),
\numberthis
\label{eq:general_master_equation_x}
\end{align*}
where the operators $\mathcal{L}_0$ and $\mathcal{L}_1$ are given by
\begin{equation}
\mathcal{L}_\sigma=\sum_{m=1}^M \left( -S_m\partial_x + \frac{{S_m}^2}{2\Omega}\partial_x^2 \right)r_{m,\sigma}(x) 
.
\label{eq:general_operator_expansion}
\end{equation}
As before we have truncated the expansion of the operators in powers of $\Omega^{-1}$ after the sub-leading terms. In explicit form we have
\begin{equation}
\mathcal{L}_\sigma=
-\partial_x v_\sigma(x)
+\frac{1}{2 \Omega}\partial_x^2 w_\sigma(x),
\label{eq:general_truncated_operators}
\end{equation}
where we have introduced
\begin{align*}
v_\sigma(x)&=\sum_{m=1}^M S_m r_{m,\sigma}(x) , \\
w_\sigma(x)&=\sum_{m=1}^M S_m^2 r_{m,\sigma}(x) .
\numberthis
\label{eq:general_VW}
\end{align*}
Again, the process described by Eqs.~\eqref{eq:general_master_equation_x} contains multiplicative noise and it is difficult to solve these equations in general. In the next section, we begin by analysing the dynamics in the limit of an infinite population.
 
\subsection{Leading-order approximation: piecewise-deterministic Markov process}
We first analyse the process in the limit of infinite system size. Here, we outline the main steps of the analysis and give the central results. The details of the calculation can be found in Appendix \ref{sec:pdmpstat}.

As before we write $\phi$ instead of $x$ in the limit of an infinite system, and we find the PDMP dynamics
\begin{align*}
\partial_t \p(\phi,0)=&
 - \partial_\phi \left[ v_0(\phi) \p(\phi,0) \right] \\
 &- \lambda_+(\phi) \p(\phi,0) + \lambda_-(\phi) \p(\phi,1), \\
\partial_t \p(\phi,1)=& 
 - \partial_\phi \left[ v_1(\phi) \p(\phi,1) \right] \\
 &+ \lambda_+(\phi) \p(\phi,0) - \lambda_-(\phi) \p(\phi,1).
\numberthis
\label{eq:general_master_equation_phi}
\end{align*}
This indicates a flow of the form
\be\label{eq:pdmp}
\dot \phi=v_{\sigma}(\phi),
\ee
in between switches of the environment. These switches in turn occur with rates $\lambda_\pm(\phi)$.

We now proceed by assuming that the deterministic flow in each of the environments is towards a unique fixed point, i.e., that there are points $\phi_\sigma^*$ for $\sigma\in\{0,1\}$ such that
\begin{align*}
v_\sigma(\phi_\sigma^*)=0.
\numberthis
\label{eq:general_fixed_points}
\end{align*}
For the time being we assume that there is only one such fixed point per state; for a more general case see Section~\ref{sec:multfp}. We assume $\phi_0^*<\phi_1^*$ without loss of generality. After a potential transient the PDMP will eventually be confined to the interval $(\phi_0^*,\phi_1^*)$. Our assumption above (only one fixed point in each environmental state) implies that the $v_\sigma$ do not change sign on this interval, $v_0(\phi)<0$ and $v_1(\phi)>0$ for $\phi\in(\phi_0^*,\phi_1^*)$: the flow is always towards fixed point $\phi_0^*$ in state $\sigma=0$, and towards $\phi_1^*$ in state $\sigma=1$, as also iillustrated in Fig.~\ref{fig:simple_fps}.

\begin{figure}
\includegraphics[width=0.45\textwidth]{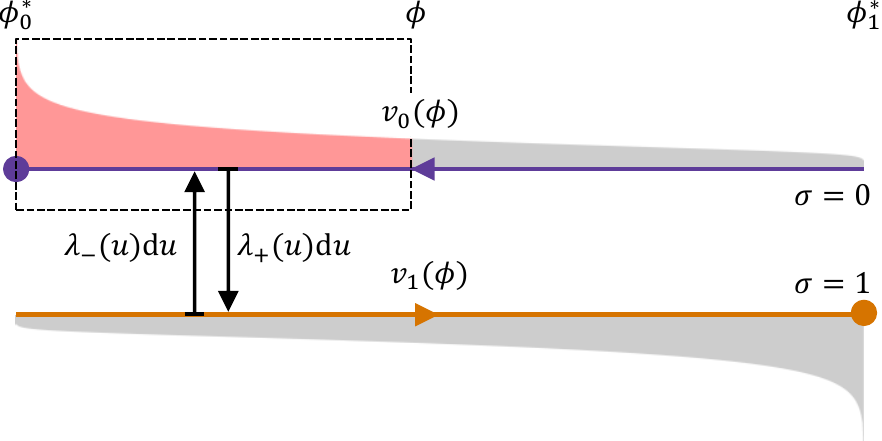}
\caption{Illustration of the physical interpretation of the currents $J_\sigma(\phi)$ (see text).}
\label{fig:currents}
\end{figure}

To analyse the PDMP further, it is useful to introduce currents
\begin{eqnarray}
J_0(\phi)&=&v_0(\phi)\Pi(\phi,0)\nonumber \\
&&-\int_{\phi_0^*}^\phi
 \left[-\lambda_+(u)\Pi(u,0)+\lambda_-(u)\Pi(u,1)\right] \d u, \nonumber \\
J_1(\phi)&=&v_1(\phi)\Pi(\phi,1)\nonumber \\
&&-\int_{\phi_0^*}^\phi \left[\lambda_+(u)\Pi(u,0)-\lambda_-(u)\Pi(u,1)\right] \d u.
\label{eq:maincurrents}
\end{eqnarray}
The physical interpretation of these currents is illustrated in Fig.~\ref{fig:currents}. 
We focus on a domain $(\phi_0^\ast, \phi)$, where $\phi_0^\ast\leq \phi\leq \phi_1^\ast$. The first term of the RHS of Eqs.~\eqref{eq:maincurrents} accounts for the probability flowing \emph{out of} such a domain at location $\phi$ due to the Liouville flow in environmental state $\sigma$. The term containing the integral describes the \emph{net flow} of probability out of the domain due to switching between the environmental states. Thus, the quantity $J_\sigma(\phi+\Delta\phi)-J_\sigma(\phi)$ represents the total amount of probability leaving the interval $(\phi, \phi+ \Delta \phi)$ per unit time.

This leads to equations of continuity
\begin{equation}
\partial_t \Pi(\phi,\sigma)=-\partial_\phi J_\sigma(\phi).
\end{equation}
In the stationary state the currents are divergence-free ($\partial_\phi J^*_\sigma=0$). Using the zero-current boundary conditions at $\phi_0^*$ and $\phi_1^*$ we find $J^*_\sigma(\phi)\equiv 0$ throughout. Summing the two stationary currents, $J^*_0(\phi)+J^*_1(\phi)=0$, we immediately find $\Pi^*(\phi,1)=-[v_0(\phi)/v_1(\phi)]\Pi^*(\phi,0)$, which allows one to replace $\Pi^*(\phi,1)$ in favour of $\Pi^*(\phi,0)$ (or vice versa) in the zero-current conditions. This results in two closed equations for $\Pi^*(\phi,0)$ and $\Pi^*(\phi,1)$ respectively. These can then be integrated directly, and we find
\begin{align*}
\Pi^*(\phi,0)&
=\frac{\mathcal{N}}{-v_0(\phi)} h(\phi),\\
\Pi^*(\phi,1)&
=\frac{\mathcal{N}}{\phantom{-}\!\!v_1(\phi)\,\,}h(\phi),
\numberthis
\label{eq:general_dist_phi}
\end{align*}
where $\mathcal{N}$ is a normalisation constant. The function $h(\phi)$ is given by 
\begin{align*}
h(\phi)\equiv \exp \left[ - \int^\phi \! \left( \frac{\lambda_+(u)}{ v_0(u)} + \frac{\lambda_-(u)}{ v_1(u)} \right) \, \d u \right].
\numberthis
\label{eq:general_dist_phi_h}
\end{align*}
Further details of the calculation are given in Appendix~\ref{sec:pdmpstat}.

Later we will need the stationary conditional probability $\Pi^*(\sigma|\phi$)---the stationary probability of having environmental state $\sigma$ given the population has state $\phi$. As discussed above we have $v_0(\phi) \Pi^*(\phi,0)+ v_1(\phi) \Pi^*(\phi,1)=0$. Writing $\Pi^*(\phi,\sigma)=\Pi^*(\phi)\Pi^*(\sigma|\phi)$ and using $\Pi^*(0|\phi)+\Pi^*(1|\phi)=1$ in Eq.~\eqref{eq:general_dist_phi} leads to
\begin{align*}
\Pi^*(0|\phi)&=\frac{ v_1(\phi)}{v_1(\phi)-v_0(\phi)} , \\
\Pi^*(1|\phi)&=\frac{-v_0(\phi)}{v_1(\phi)-v_0(\phi)} .
\numberthis
\label{eq:app_state_given_phi}
\end{align*}
It is perhaps surprising that these conditional probabilities are independent of the switching rates $\lambda_+$ and $\lambda_-$.

\subsection{Sub-leading order: linear-noise approximation}
We now proceed by including contributions of intrinsic noise to sub-leading order. We focus on a time interval between switches of the environment, i.e., we assume $\sigma$ is constant. During such time intervals the environmental noise has no effect, and the problem reduces to that of a conventional individual-based system with a fixed environment. Following established procedures we write $n/\Omega=\phi(t)+\Omega^{-1/2}\xi(t)$. The deterministic dynamics is given by $\dot \phi=v_\sigma(\phi)$. The outcome of a standard LNA can be expressed as a linear Langevin equation
\begin{equation}
\dot \xi(t)=v_\sigma'(\phi)\xi+\sqrt{w_\sigma(\phi)}\eta(t)
\label{eq:lna}
\end{equation}
for fluctuations about $\phi(t)$, where 
\begin{equation}
w_\sigma(\phi)=\sum_m S_m^2 r_{m,\sigma}(\phi).
\end{equation}
These relations describe the evolution of the population (within the LNA) between switches of the environment. When a change of the environment occurs the variable $\sigma(t)$ changes at a discrete point in time, and the next such interval of constant $\sigma$ begins.

Within the LNA the evolution of the probability to observe state $\phi,\xi,\sigma$ is described by the following set of equations
\begin{widetext}
\begin{align}
\partial_t \Pi(\phi,\xi,0)= {}&
-\partial_\phi \left[ v_0(\phi) \Pi(\phi,\xi,0)\right]- v_0^\prime(\phi)\, \partial_\xi \left[\xi \Pi(\phi,\xi,0)\right] + \frac{w_0(\phi)}{2} \partial_\xi^2 \Pi(\phi,\xi,0)
- \lambda_+ ( \phi ) \Pi(\phi,\xi,0)+ \lambda_- ( \phi ) \Pi(\phi,\xi,1), \nonumber \\
\partial_t \Pi(\phi,\xi,1) ={}&
-\partial_\phi \left[ v_1(\phi) \Pi(\phi,\xi,1)\right]- v_1^\prime(\phi)\, \partial_\xi \left[\xi \Pi(\phi,\xi,1)\right] 
+ \frac{w_1(\phi)}{2} \partial_\xi^2 \Pi(\phi,\xi,1)+ \lambda_+ ( \phi ) \Pi(\phi,\xi,0)
- \lambda_- ( \phi ) \Pi(\phi,\xi,1).
\label{eq:masterxi}
\end{align}
\end{widetext}
While the expressions on the right-hand side look complicated, the different terms have a clear physical meaning. The Liouvillian terms $-\partial_\phi[v_\sigma(\phi)\Pi(\phi,\xi,\sigma)]$ describe the deterministic flow [$\dot \phi=v_\sigma(\phi)$] between switches of the environment. The Fokker--Planck-like terms $-v_\sigma^\prime(\phi) \partial_\xi\left[ \xi \Pi(\phi,\xi,\sigma) \right]
+ w_\sigma(\phi) \partial_\xi^2 \Pi(\phi,\xi,\sigma)/2$ capture the evolution of $\xi$ within the LNA of Eq.~\eqref{eq:lna}. Finally, the terms proportional to $\lambda_\pm(\phi)$ describe switching of the environment. Consistent with the expansion in the system-size $\Omega$ we have replaced $\lambda_\pm(x)$ by $\lambda_\pm(\phi)$, i.e., any dependence of the switching rates on the state of the population is taken to be on the state of the PDMP $\phi$. 

\subsection{Stationary state within the linear-noise approximation}
At stationarity the time derivatives on the LHS of Eqs.~\eqref{eq:masterxi} vanish. Using asterisks as before to denote stationary distributions, and writing $\Pi^*(\phi,\xi,\sigma)=\Pi^*(\phi,\sigma)\Pi^*(\xi|\phi,\sigma)$, we find 
\begin{widetext}
\begin{align}
0 = {}& - \Pi^*(\phi,0)\, v_0^\prime(\phi)\, \partial_\xi \left[\xi \Pi(\xi|\phi,0)\right] 
+\frac{1}{2} \Pi^*(\phi,0)\,w_0(\phi)\, \partial_\xi^2 \Pi^*(\xi|\phi,0) -\partial_\phi \left[
v_0(\phi) \,\Pi^*(\phi,0)\, \Pi^*(\xi|\phi,0) \right] \nonumber \\
{}& - \Pi^*(\phi,1)\, v_1^\prime(\phi)\, \partial_\xi \left[\xi \Pi(\xi|\phi,1)\right] 
+\frac{1}{2} \Pi^*(\phi,1)\,w_1(\phi)\, \partial_\xi^2 \Pi^*(\xi|\phi,1) -\partial_\phi \left[
v_1(\phi) \,\Pi^*(\phi,1)\, \Pi^*(\xi|\phi,1) \right]
\end{align}
\end{widetext}
from summing the two equations~\eqref{eq:masterxi} at stationarity. 

At this point we introduce a further approximation. We assume $\Pi^*(\xi|\phi,0)\approx \Pi^*(\xi|\phi,1)$, and simply write $\Pi^*(\xi|\phi)$ for either of these. This will be justified below. Making this assumption leads to
\BE
0&=&
- \Big( \Pi^*(0|\phi)v_0^\prime(\phi) 
+ \Pi^*(1|\phi)v_1^\prime(\phi) \Big) \frac{\partial}{\partial \xi}\xi\Pi^*(\xi|\phi)\nonumber \\
&&+ \frac{1}{2} \Big(
\Pi^*(0|\phi)w_0(\phi) + \Pi^*(1|\phi)w_1(\phi) \Big) \frac{\partial^2}{\partial \xi^2} \Pi^*(\xi|\phi),\nonumber \\
\label{eq:helpme}
\EE
where we have used the relation $v_0(\phi) \Pi^*(\phi,0)+ v_1(\phi) \Pi^*(\phi,1)=0$, valid at stationarity, to show that the 
terms containing derivatives with respect to $\phi$ cancel out.

Eq.~(\ref{eq:helpme}) resembles a stationary Fokker--Planck equation. The drift and diffusion coefficients are obtained from the $v_\sigma^\prime$ and $w_\sigma$, weighted by the likelihood to find the environment in each of its states when the PDMP is at $\phi$.

Solving the stationary equation (\ref{eq:helpme}) is standard \cite{risken1984fokker}; we find a Gaussian distribution, $\Pi^*(\xi|\phi)$, with zero mean and with variance 
\be
s^2(\phi)=-\frac{1}{2}\frac
{
\Pi^*(0|\phi) w_0(\phi)
+\Pi^*(1|\phi) w_1(\phi)
}
{
 \Pi^*(0|\phi) v^\prime_0(\phi)
+\Pi^*(1|\phi) v^\prime_1(\phi)
}
.
\label{eq:app_noise_variance}
\ee
Using Eq.~\eqref{eq:app_state_given_phi} this can be simplified to give
\begin{align*}
s^2(\phi)=\frac{1}{2}\frac
{w_0(\phi)v_1(\phi)-w_1(\phi)v_0(\phi)}{v_0(\phi)v_1^\prime(\phi) - v_1(\phi)v_0^\prime(\phi)}
.
\numberthis
\label{eq:app_noise_variance2}
\end{align*}

\begin{figure}
\captionsetup[subfigure]{labelformat=empty}
{\centering
\includegraphics[width=0.96\columnwidth]{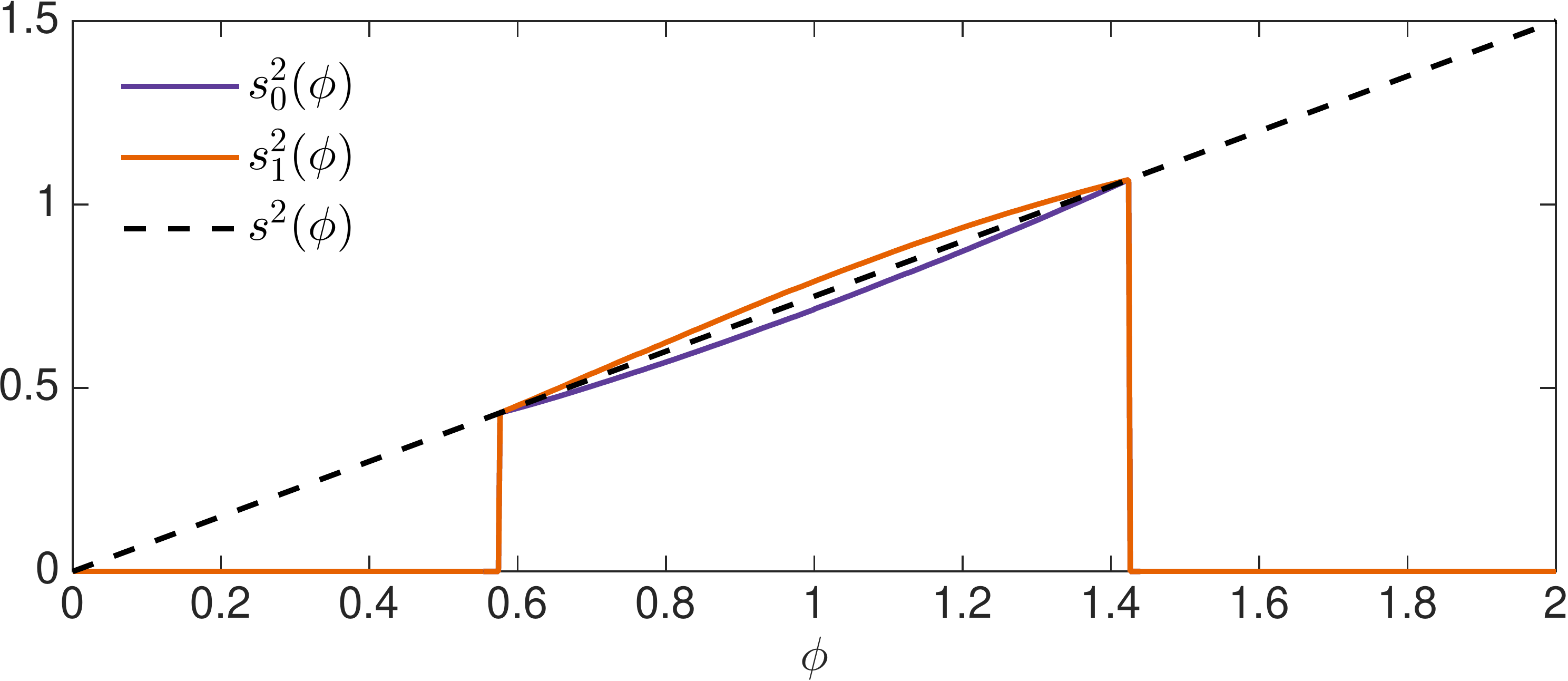}}
\caption{The variance of $\xi$, conditioned on the state $\phi$ of the PDMP and the state of the environment (see text for details). Results are shown for the dimerisation process of Section~\ref{subsec:dimer}. Solid lines are $s_0^2(\phi)$ and $s_1^2(\phi)$ obtained from simulations (see text for definitions). The dashed line shows the approximation of Eq.~\eqref{eq:app_noise_variance2}. Parameters used are $b_0=0.667$, $b_1=4.048$, $d=2$, $\lambda_+=\lambda_-=1$ and $\Omega=150$.}
\label{fig:variance_at_phi}
\end{figure}

We conclude this section with a brief comment on the approximation $\Pi^*(\xi|\phi,\sigma)\approx \Pi^*(\xi|\phi)$. In the fast-switching limit the approximation is plausible, the system switches between environmental states too fast for the variable $\xi$ to equilibrate to a distribution specific to the environmental state. For this case, the approximation reproduces the result of \cite{thomas2014phenotypic}. In the limit of slow switching on the other hand the system spends most of its time near the fixed points $\phi_0^*$ and $\phi_1^*$. Our approach then also recovers appropriate distributions of $\xi$. For example, when $\phi\approx\phi_0^*$ we have $v_0(\phi=\phi_0^*)=0$ and Eq.~(\ref{eq:app_noise_variance2}) hence reduces to $s^2(\phi_0^*)=-[w_0(\phi_0^*)/(2v_0'(\phi_0^*))]$. This is the stationary variance of the process $\dot \xi=-v_0'(\phi_0^*)\xi+\sqrt{w_0(\phi_0^*)}\eta$. A similar argument applies when $\phi\approx\phi_1^*$. Outside the limits of fast and slow switching, the quality of the approximation can be verified numerically. In Fig.~\ref{fig:variance_at_phi} we illustrate this for a model with nonlinear dynamics (discussed in more detail in the next section). To test the validity of the above assumption we have implemented the following measurement protocol: We first generate a combined realisation of the environmental process and the PDMP, i.e., a realisation of the process defined by Eqs.~(\ref{eq:general_master_equation_phi}). Subsequently we generate a realisation of the birth-death dynamics in a finite population for the same path of the environment. In the stationary state we then measure the variance of $\xi$ conditioned on the state $\phi$ of the PDMP and on the environmental state $\sigma$. We denote this variance by $s^2_\sigma(\phi)$. This is then averaged over multiple realisations, and compared against the approximation of Eq.~\eqref{eq:app_noise_variance2}. The data in Fig.~\ref{fig:variance_at_phi} shows that the approximation works well, with only slight deviations in the region away from the two fixed points $\phi_0^*$ and $\phi_1^*$. For the linear model numerical results and the theoretical prediction are virtually indistinguishable.

\section{Further Examples}
\label{sec:examples}
\subsection{Nonlinear reactions rates}
\label{subsec:dimer}
In order to demonstrate the generality of our approach, we now proceed to a model system with nonlinear reaction rates. The system is identical to the one described in Eq.~\eqref{eq:simple_reactions_all}, except that the last reaction (removal of individuals) is replaced by
\begin{align}
2\mathcal{A} \xrightarrow{\mathmakebox[0.5cm]{d/\Omega}} \varnothing. \label{eq:nonlinear}
\end{align}
The above notation indicates that this last reaction occurs with rate $dn(n-1)/(2\Omega)$, where $n$ is the number of individuals in the system. The reaction can be interpreted as a dimerisation process, in which two particles of type $\mathcal{A}$ form a complex which is chemically inert and hence effectively removed. In addition to the switches between $\mathcal{G}_0$ and $\mathcal{G}_1$, this system is described by two reactions, with stoichiometric coefficients and reaction rates
\begin{align}
S_1={}&+1, \ r_{1,\sigma}=b_\sigma, \nonumber \\
S_2={}&-2, \ r_{2,\sigma}=\frac{d}{2} x^2.
\end{align}
We have written $x=n/\Omega$ as before.
Using the notation of the preceding sections, we have the following drift and the diffusion terms
\begin{align}
v_\sigma(x)= b_\sigma- x^2d, \quad \text{and} \quad 
w_\sigma(x)=b_\sigma+2x^2d .
\label{eq:dimer_VW}
\end{align}
In the limit $\Omega\to\infty$, we obtain a PDMP with fixed points $\phi_\sigma^*=\sqrt{b_\sigma/d}$ for $\sigma=0,1$. As before we assume $b_0<b_1$. The stationary distribution of the PDMP is found from Eq.~\eqref{eq:general_dist_phi}, 
\begin{align*}
\Pi^*(\phi,0)&
=\frac{\mathcal{N}}{\phi^2-{\phi_0^*}^2}
\left(
\frac{\phi-\phi_0^*}{\phi_0^*+\phi}
\right)^{\frac{\lambda_+}{\sqrt{4 b_0 d}}}
\left(
\frac{\phi_1^*-\phi}{\phi_1^*+\phi}
\right)^{\frac{\lambda_-}{\sqrt{4 b_1 d}}}
, \\
\Pi^*(\phi,1)&
=\frac{\mathcal{N}}{ {\phi_1^*}^2 - \phi^2 }
\left(
\frac{\phi-\phi_0^*}{\phi_0^*+\phi}
\right)^{\frac{\lambda_+}{\sqrt{4 b_0 d}}}
\left(
\frac{\phi_1^*-\phi}{\phi_1^*+\phi}
\right)^{\frac{\lambda_-}{\sqrt{4 b_1 d}}}
,
\numberthis
\label{eq:dimer_dist_phi}
\end{align*}
for $\phi\in(\phi_0^*,\phi_1^*)$, and where $\mathcal{N}$ is the usual normalisation constant.

\begin{figure}
\captionsetup[subfigure]{labelformat=empty}
{\centering
\includegraphics[width=\columnwidth]{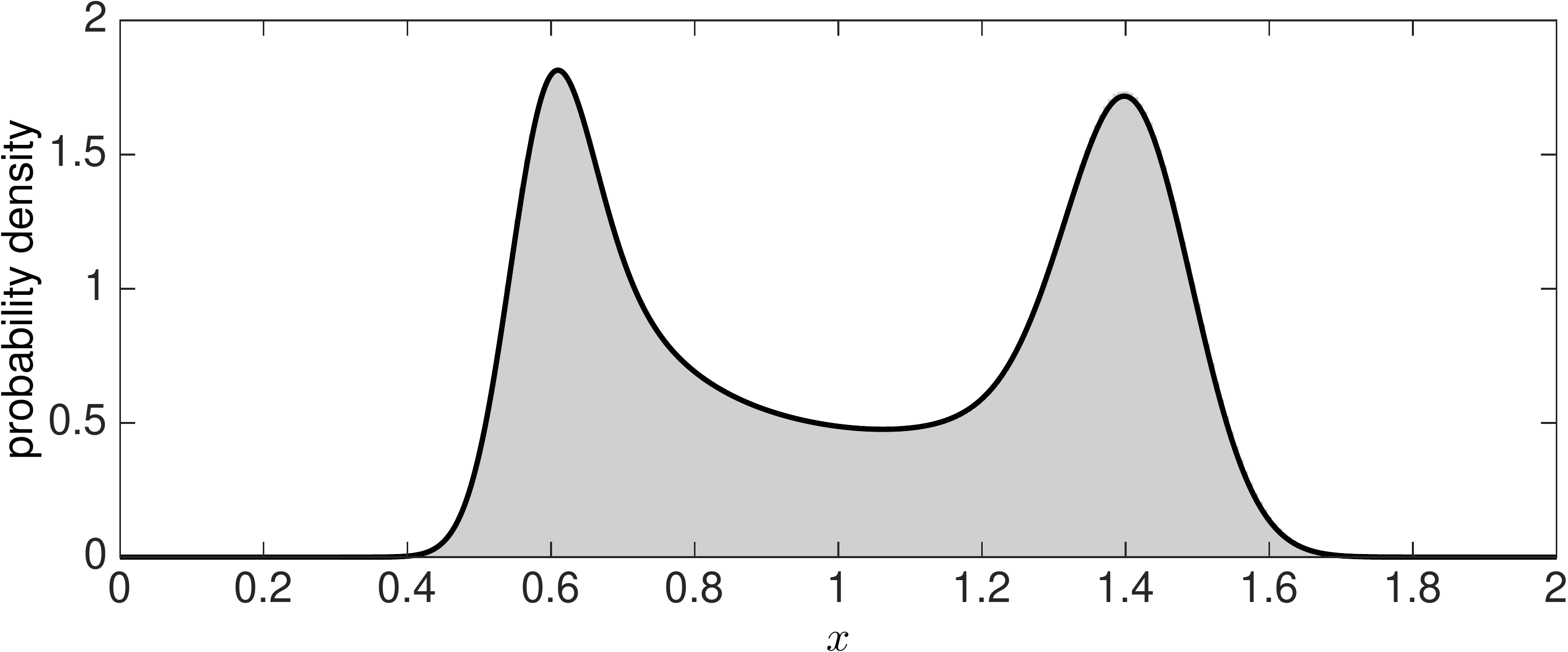}}
\caption{Stationary distribution for the nonlinear model of Sec. \ref{subsec:dimer}. The solid line is the theoretical prediction, and the histogram is data obtained from simulations using the Gillespie algorithm. Parameters are $b_0=0.667$, $b_1=4.048$, $d=2$, $\lambda_+=\lambda_-=1$ and $\Omega=150$.}
\label{fig:dimer}
\end{figure}

From Eq.~\eqref{eq:app_noise_variance2} finally we find $s^2 (\phi) = 3 \phi / 4$, i.e.,
\begin{align}
\Pi^*(\xi|\phi)=\sqrt{\frac{2}{3\pi\phi}}\exp\left(-\frac{2}{3\phi}\xi^2\right).
\end{align}
The stationary distribution $\Pi^*(x)$ is then obtained by numerically evaluating Eq.~\eqref{eq:full}. Results are compared against simulation in Fig.~\ref{fig:dimer}, and we find convincing agreement between theoretical predictions and simulations. This confirms the validity of the assumptions and approximations made during the course of the analytical calculation.

\subsection{System-dependent environmental transition rates}
\label{subsec:feedback}
We now turn to another variation of the original linear model [Eqs.~\eqref{eq:simple_reactions_all}]. For the dynamics within the population we use the same reactions and rates as in Eqs.~\eqref{eq:simple_reactions_all}, but we consider the case in which the rates with which the environment switches between states depends on the state of the population, i.e.,
\BE
\mathcal{G}_0 &\xrightarrow{\lambda_+ (n)} &\mathcal{G}_1, \nonumber \\
\mathcal{G}_1 &\xrightarrow{\lambda_- (n)}&\mathcal{G}_0, 
\label{eq:feedback_environmental_switching}
\EE
where we choose the linear form $\lambda_\pm (n)= \alpha_\pm + \beta_\pm (n / \Omega)$. The coefficients $\alpha_\pm$ and $\beta_\pm$ are constants, chosen such that $\lambda_\pm$ remain non-negative. 

The drift and diffusion terms of the dynamics within the population are 
\begin{align}
v_\sigma(x)=b_\sigma-xd,\quad \text{and} \quad w_\sigma(x)=b_\sigma+xd .
\label{eq:feedback_VW}
\end{align}
as before. The stationary distribution of the PDMP in the limit of infinite populations is obtained from Eqs.~\eqref{eq:general_dist_phi} as
\begin{align*}
&\Pi^*(\phi,0)= 
\mathcal{N}
e^{\frac{\beta_+ + \beta_- }{d} \phi }
\frac{
\left( \phi - \phi^*_0 \right)^{\kappa_+}
\left( \phi^*_1 - \phi \right)^{\kappa_-}
}
{\phi-\phi^*_0}
, \\
&\Pi^*(\phi,1)=\mathcal{N}
e^{\frac{\beta_+ + \beta_- }{d} \phi}
\frac{
\left( \phi - \phi^*_0 \right)^{\kappa_+}
\left( \phi^*_1 - \phi \right)^{\kappa_-}
}
{\phi^*_1 - \phi}
,
\numberthis
\label{eq:feedback_dist_phi}
\end{align*}
where the exponents $\kappa_\pm$ are given by 
\begin{align}
\kappa_\pm=\frac{\alpha_\pm}{d}+\frac{\beta_\pm b_0}{d^2}.
\label{eq:feedback_exponents}
\end{align}
The dynamics within the population is the same as in the linear model above, so evaluating Eq.~\eqref{eq:app_noise_variance2} again leads to $s^2(\phi)=\phi$. 
The theoretical estimate of the stationary distribution $\Pi^*(x)$, finally, is again found from numerical integration of Eq.~\eqref{eq:full}.
\begin{figure}
\captionsetup[subfigure]{labelformat=empty}
{\centering\
\subfloat[]{\includegraphics[width=\columnwidth]{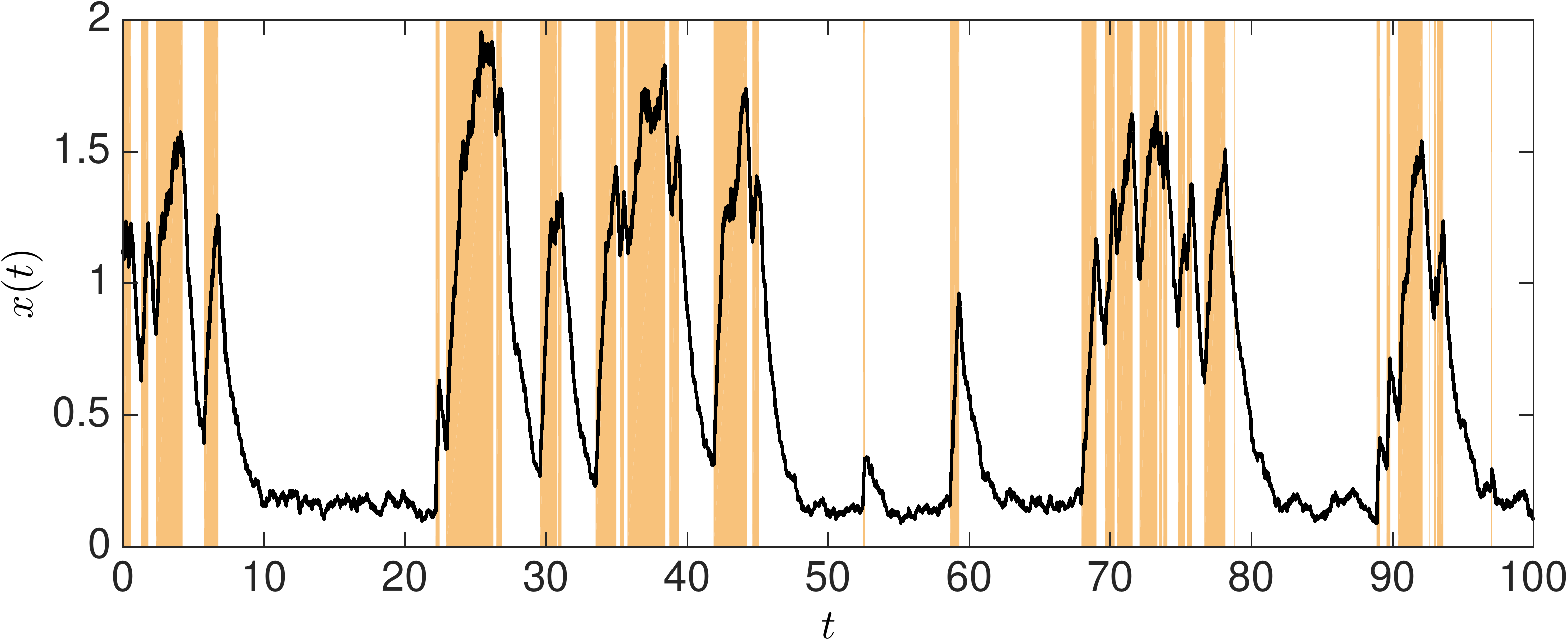}} \\
\subfloat[]{\includegraphics[width=\columnwidth]{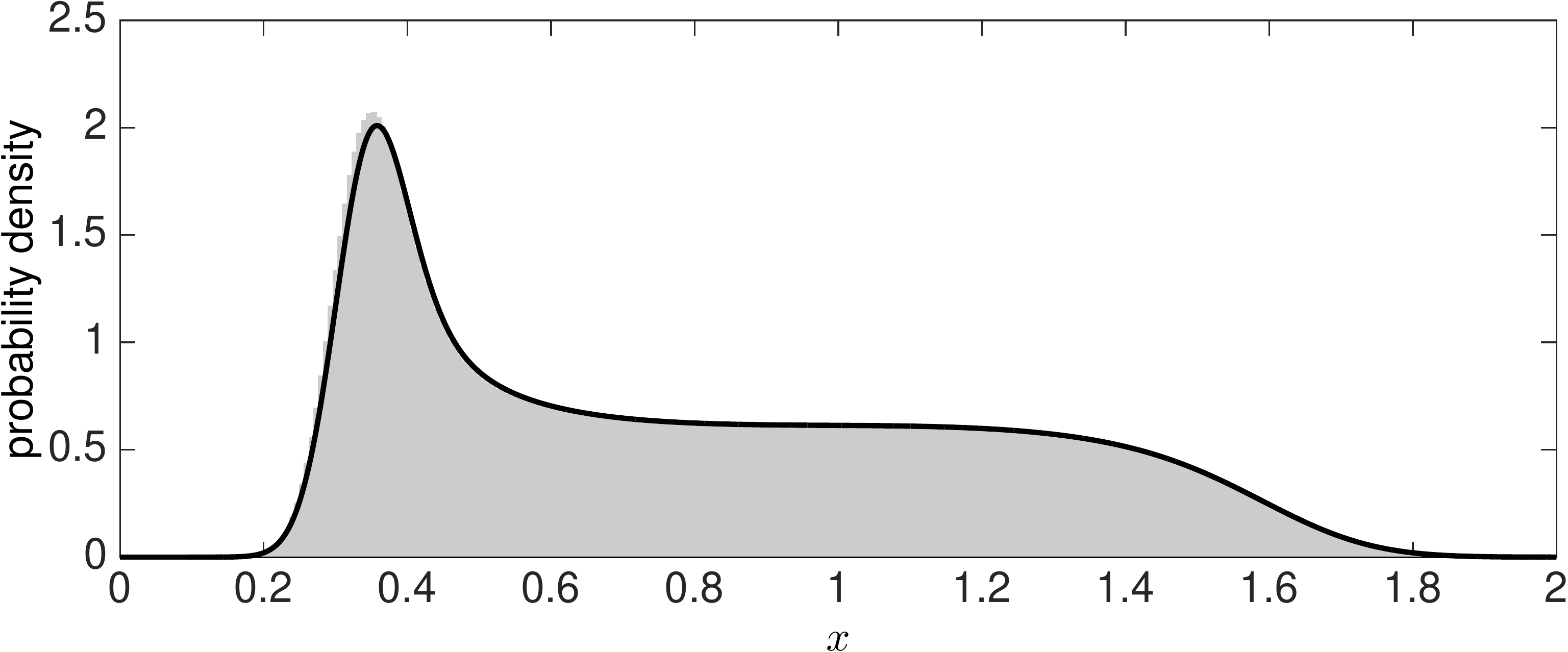}} }
\caption{Model in which rates of environmental switching depends on the state of the population. The upper panel shows a typical time course, as generated by the Gillespie algorithm. The lower panel depicts the stationary distribution, the histogram is from simulations, and the solid line represents the approximation of our theory. Model parameters are $\alpha_\pm=0, \beta_\pm=1, b_0=1/3, b_1=5/3, d=1$ and $\Omega=150$.}
\label{fig:feedback}
\end{figure}

Fig.~\ref{fig:feedback} shows a sample path of the dynamics and the stationary distribution for the choice $\lambda_+(x)=\lambda_-(x)=x$ (i.e., $\alpha_\pm=0$ and $\beta_\pm=1$). In this case, transitions between states are more likely at higher concentrations of the system. Again, we find good agreement between the predicted distribution and the simulation results. It is important to note that the parameters we used in the figure are not special in any way: we have tested other choices of the parameter, and we find an agreement between simulations and theory of a similar quality.

\subsection{Multiple fixed points}
\label{sec:multfp}
In the final example, we consider more complicated dynamics within the population such that there are multiple fixed points of the flows $v_\sigma(x)$. The switching between the two environmental states is taken to occur with constant rates $\lambda_\pm$. Specifically, the population dynamics is now modelled by the following reactions
\begin{align}
 \mathcal{G}_\sigma \xrightarrow{\mathmakebox[0.97cm]{\vphantom{1^1}\Omega c_{1,\sigma}}}{}& \mathcal{G}_\sigma + \mathcal{A}, \nonumber \\
 \mathcal{G}_\sigma+\mathcal{A} \xrightarrow{\mathmakebox[0.97cm]{\vphantom{1^1} c_{2,\sigma}}}{}& \mathcal{G}_\sigma, \nonumber \\
 \mathcal{G}_\sigma +2\mathcal{A} \xrightarrow{\mathmakebox[0.97cm]{\vphantom{1^1} c_{3,\sigma}/\Omega}}{}& \mathcal{G}_\sigma + 3\mathcal{A}, \nonumber \\
 \mathcal{G}_\sigma +3\mathcal{A} \xrightarrow{\mathmakebox[0.97cm]{\vphantom{1^1} c_{4,\sigma}/\Omega^2}}{}& \mathcal{G}_\sigma + 2\mathcal{A}, 
\end{align}
with constant parameters $c_{i,\sigma}>0$.
The corresponding stoichiometric coefficients for the four reactions are
\begin{align}
S_1=S_3=+1,\ S_2=S_4=-1, 
\end{align}
and the propensities $r_{m,\sigma}(x)$ read
\begin{align*}
r_{1,\sigma}(x)={}&c_{1,\sigma}, && \ r_{2,\sigma}(x)=c_{2,\sigma}x, \\
r_{3,\sigma}(x)={}& \frac{c_{3,\sigma}}{2}x^2, && \ r_{4,\sigma}(x)=\frac{c_{4,\sigma}}{6} x^3. 
\numberthis
\end{align*}
Additional details of the model and the numerical values of the parameters we used for our analysis can be found in Appendix~\ref{app:multfp}.

\begin{figure}
\includegraphics[width=0.47\textwidth]{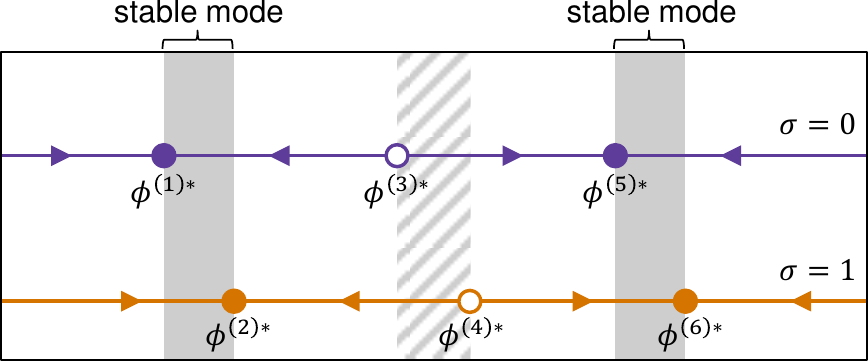}
\caption{Illustration of the Liouville flow in the two environmental states $\sigma=0,1$ for the model in Sec. \ref{sec:multfp}. Stable fixed points are shown as filled circles, and unstable fixed points as open circles. The shaded areas represent the two stable dynamic modes described in the text.}
\label{fig:multiple_fps}
\end{figure}
\begin{figure}
\captionsetup[subfigure]{labelformat=empty}
{\centering
\subfloat[]{\includegraphics[width=0.5\columnwidth]{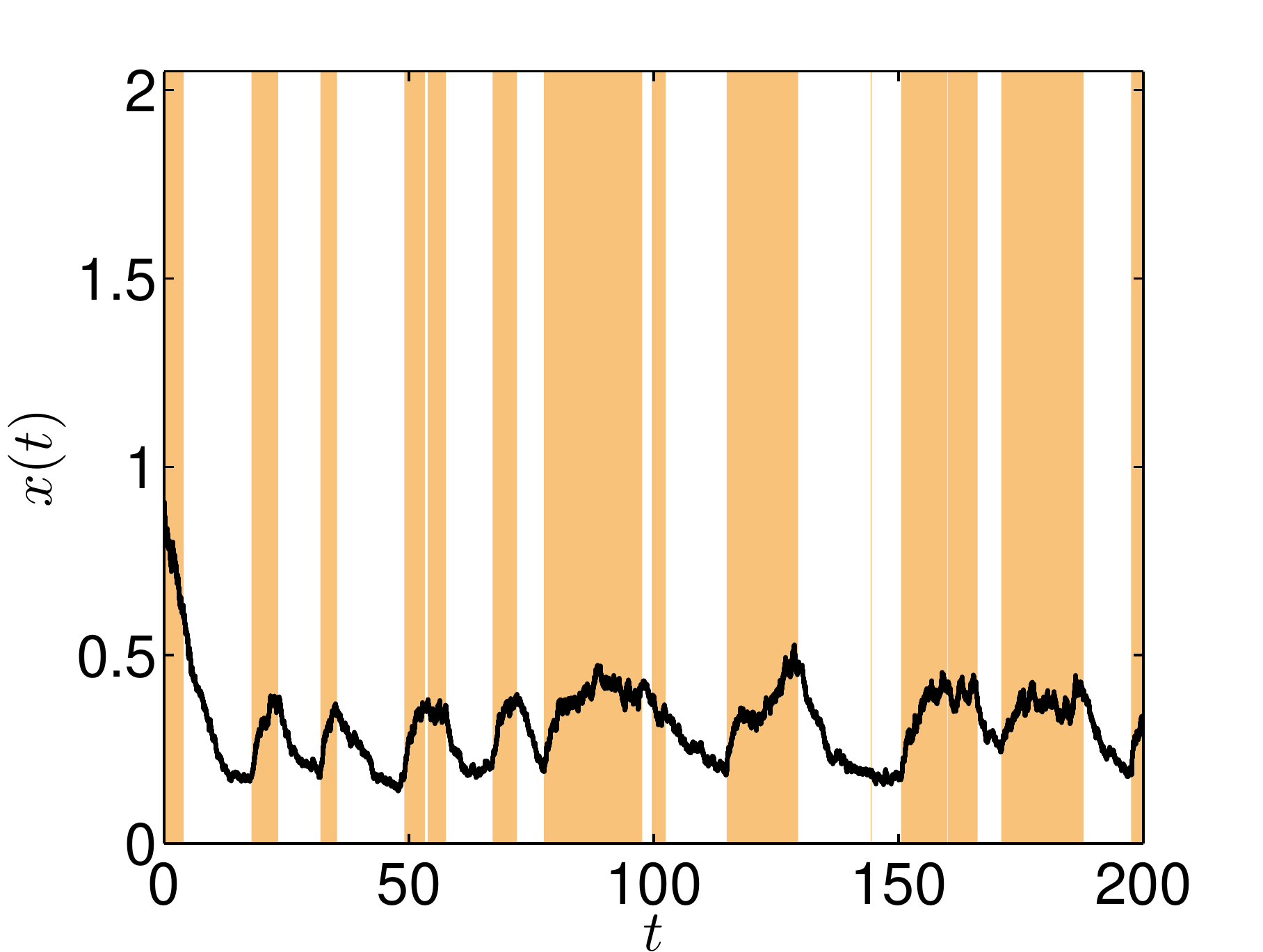}} ~
\subfloat[]{\includegraphics[width=0.5\columnwidth]{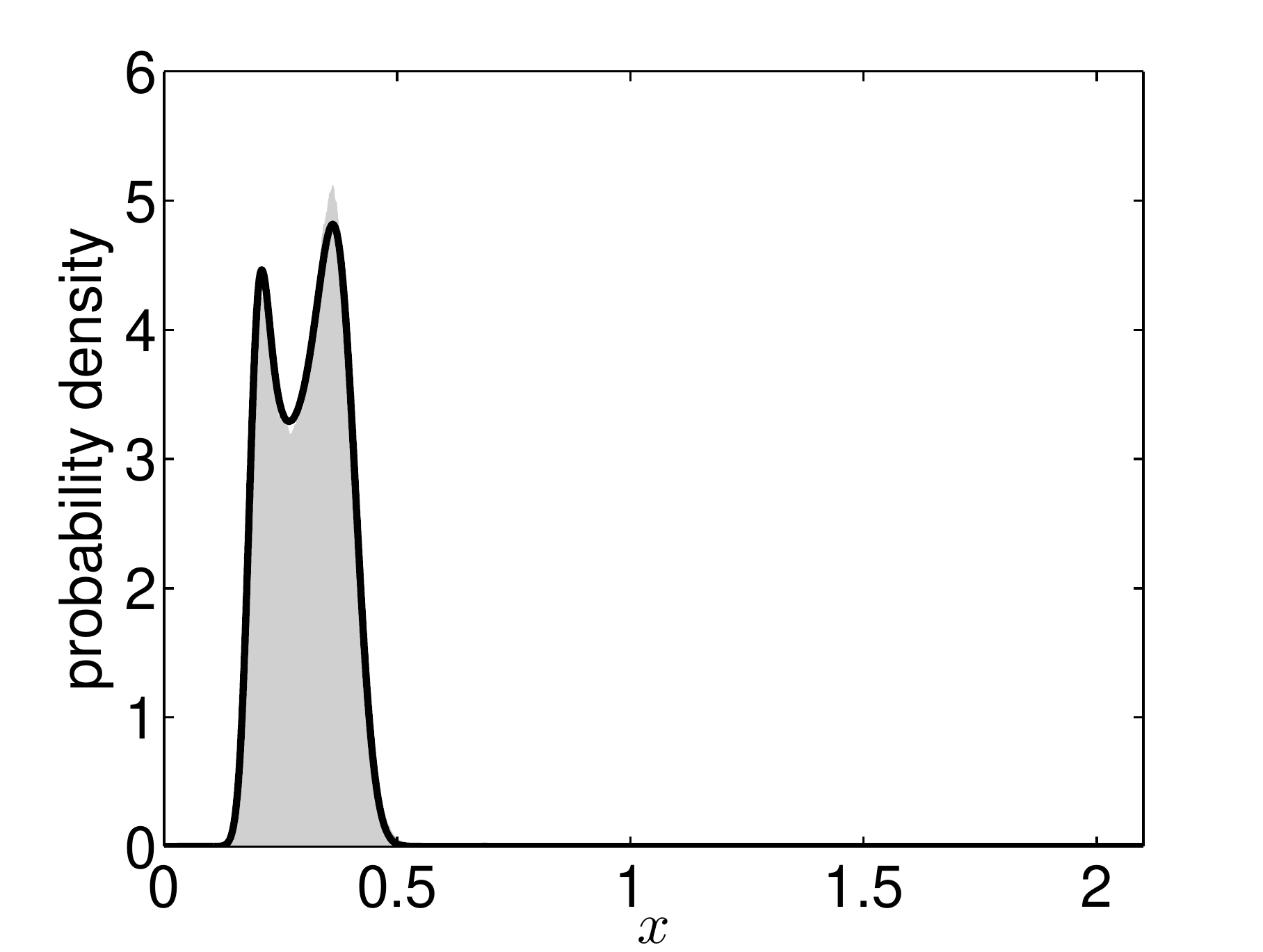}} \\
\subfloat[]{\includegraphics[width=0.5\columnwidth]{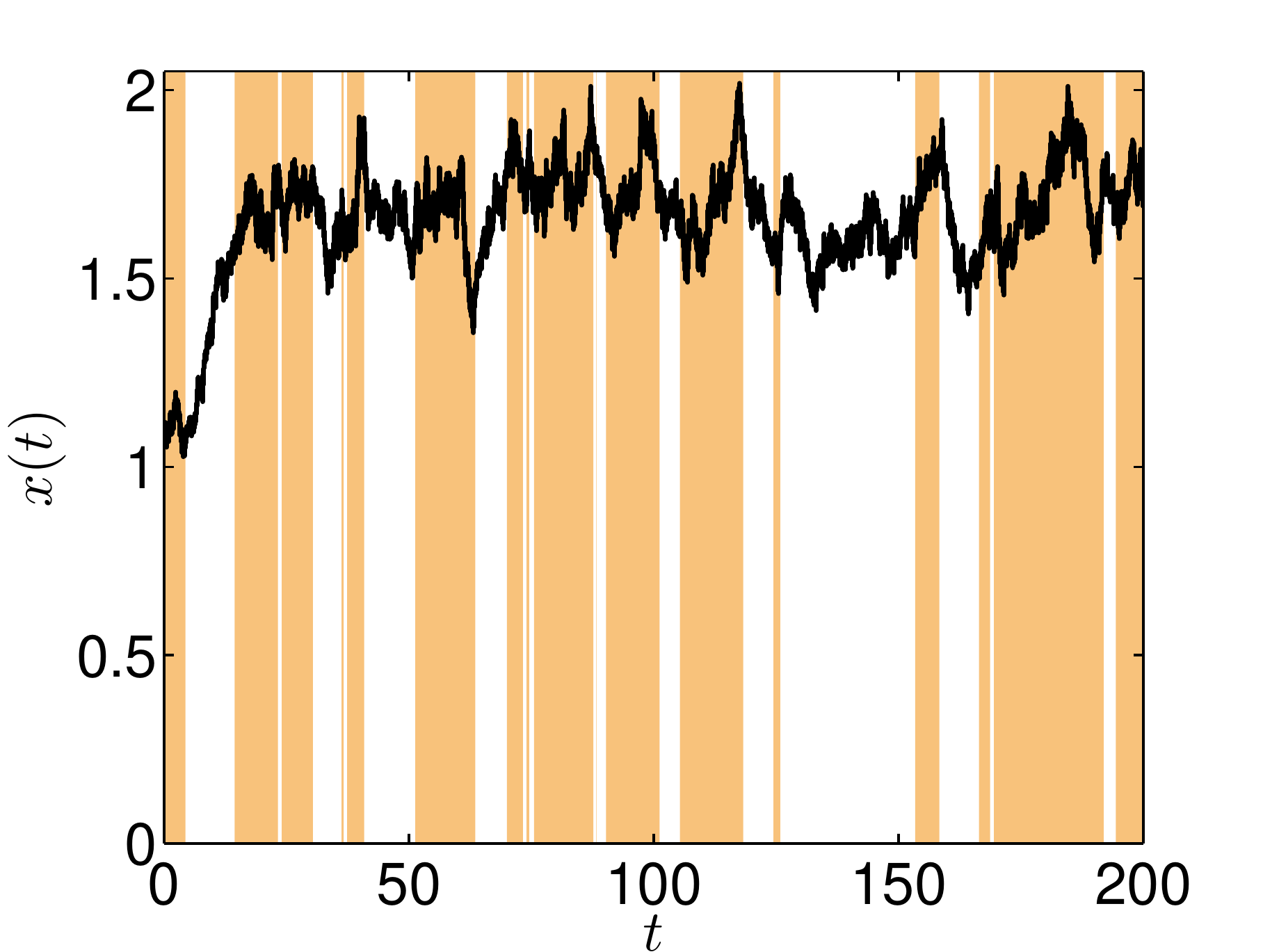}} ~
\subfloat[]{\includegraphics[width=0.5\columnwidth]{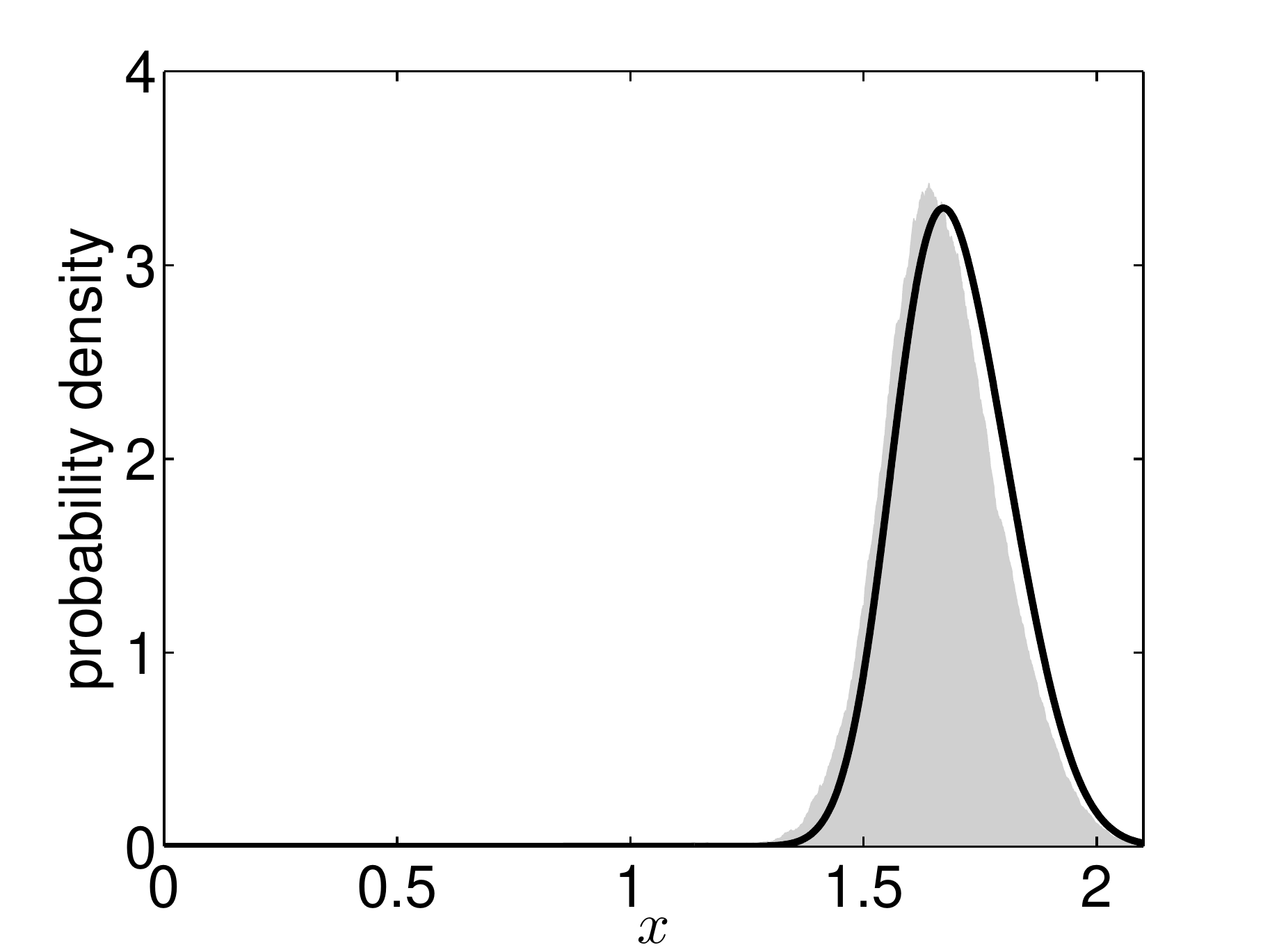}} }
\caption[Many]{Representative realisations and stationary distributions for the model with described in Sec. \ref{sec:multfp}. The upper two panels represent one dynamic mode, the lower two panels the other. More precisely, the upper panels show the outcome when the initial condition is to the left of $\phi^{(3)*}$, the lower two panels show a situation in which the dynamics is started to the right of $\phi^{(4)*}$. The histograms in panels on the right are from Gillespie simulations, and the solid lines are from the analytical approximation. Model parameters are $\lambda_+=\lambda_-=0.2$ and $\Omega=1000$, remaining parameters as described in Appendix \ref{app:multfp}.}
\label{fig:multiple_figs}
\end{figure}
Again, we first consider the PDMP, i.e., the limit $\Omega\to\infty$. For the parameters chosen for our analysis one finds three fixed points of the dynamics $\dot \phi=v_0(\phi)$, at $\phi^*\approx 0.20, 0.90, $ and $ 1.6$. We label these $\phi^{(1)*}, \phi^{(3)*}$, and $\phi^{(5)*}$. These fixed points are linearly stable, unstable, and stable respectively. In the environmental state $\sigma=1$ we have fixed points $\phi^*\approx 0.4, 1.1$ and $1.8$, labelled $\phi^{(2)*}, \phi^{(4)*}$ ,and $\phi^{(6)*}$(again stable, unstable, and stable respectively). This arrangement of fixed points is illustrated in Fig~\ref{fig:multiple_fps}. The dynamics of the PDMP depends on the initial condition. If started between the fixed points $\phi^{(1)*}$ and $\phi^{(2)*}$, the system will be confined between these two values and follow a dynamics similar to that of the system in Sec.~\ref{subsec:dimer}. Similarly, if the initial condition is between the two fixed points $\phi^{(5)*}$ and $\phi^{(6)*}$, the PDMP will operate in the interval between these two points. We will refer to these as the two `stable modes' of the PDMP dynamics. For other initial conditions the PDMP will eventually reach one of these two stable modes as well, which one this is will depend on the starting point and on the exact path the environment takes.

If the system is finite, the dynamics are subject to intrinsic noise. Two representative trajectories are shown Fig.~\ref{fig:multiple_figs}. Depending on initial conditions the system either remains close to the interval between $\phi^{(1)*}$ and $\phi^{(2)*}$ or near the interval between $\phi^{(5)*}$ and $\phi^{(6)*}$. Intrinsic noise will allow for excursions outside these intervals, similar to what was observed in Sec.~\ref{subsec:dimer}. The finite system can traverse the region between $\phi^{(2)*}$ and $\phi^{(5)*}$, at least in principle, and move from one of the dynamic modes to the other. This is similar to escaping from a basin of attraction in systems with constant environment. We generally expect that the rate with which this happens is exponentially suppressed in the noise strength. We will assume that such events are sufficiently rare so that they can be ignored for the purposes of our analysis.

The general approach we have developed can then be applied to the system in either of the two dynamic modes. Due to the complexity of the flow fields $v_\sigma(x)$, the relevant equations have to be evaluated numerically. In Fig.~\ref{fig:multiple_figs} we show the resulting predictions for the stationary distribution in either mode. As in the previous examples, comparison with data from numerical simulations shows very good agreement.

\section{Conclusion}
\label{sec:conclusion}
In summary, we have constructed a systematic approach with which to investigate the effects of demographic noise in systems subject to sudden random switches of reaction rates. We have focused on relatively simple birth-death processes of one single species and in which birth and death rates depend on both the state of the population and on the state of an external environment. The states of the environment follow a telegraph process, with transition rates which may depend on the state of the population. 
Previously existing approaches either disregard intrinsic fluctuations and only account for environmental noise, or they focus on cases in which there is a clear separation of time scales between the population dynamics and the dynamics of the environment. The former approach in particular leads to the well-established picture of so-called piecewise deterministic Markov processes. Our work systematically improves on this view; we take into account demographic fluctuations to leading order and carry out a system-size expansion and linear-noise approximation about the PDMP dynamics. 

Using the linear-noise approximation, and retaining the discreteness of the environmental process, we then approximate the resulting stationary distribution of the population dynamics. We have tested the resulting theory on a number of model systems, both with linear and nonlinear reaction rates, situations in which environmental switching depends on the state of the population, and covering systems with single dynamic modes and cases where there are multiple attractors. In all cases we have tested, the approximation is in good agreement with results from simulations. In particular no separation of time scales is required.

The technique we provide makes our understanding of processes involving both intrinsic and extrinsic noise more complete. While the existing PDMP description has been shown to be successful in many instances, it disregards intrinsic noise entirely. We are now in a position to describe the effects of demographic noise in systems with switching environments in the spirit of the van Kampen expansion. This allows us to investigate models subject to combinations of intrinsic and extrinsic noise, and in particular systems in which some degrees of freedom can be treated within a linear-noise approximation, while other variables remain fundamentally discrete. Such systems can be of relevance for example in the context of genetic switches, bacterial populations subject to varying external conditions, or to predator-prey dynamics. In order to make the method more applicable several extensions can be considered in future work. This includes the case of multiple environmental states, systems with more than one species, or indeed environmental dynamics beyond simple telegraph processes. Work along these lines is currently in progress.

\begin{acknowledgements}
PGH, YTL and TG thank the Engineering and Physical Sciences Research Council (EPSRC) for funding (grant reference EP/K037145/1).
\end{acknowledgements}

\appendix
\section{The stationary distribution of the PDMP}\label{sec:pdmpstat}
In this section we briefly discuss the calculation of the stationary distribution of the PDMP. Following on from Eq.~(\ref{eq:maincurrents}), and writing $\partial_\phi J^*_\sigma(\phi)\equiv 0$ in the stationary state, we have $J^*_\sigma(\phi)\equiv 0$ throughout, using zero-current conditions at the boundaries. This leads to 
\begin{align*}
J^*_0(\phi)={}&v_0(\phi)\Pi^*(\phi,0) \\
&-\int_{\phi_0^*}^\phi
\left[-\lambda_+(u)\Pi^*(u,0)+\lambda_-(u)\Pi^*(u,1)\right]\,\d u=0, \\
J^*_1(\phi)={}&v_1(\phi)\Pi^*(\phi,1) \\
&-\int_{\phi_0^*}^\phi \left[\lambda_+(u)\Pi^*(u,0)-\lambda_-(u)\Pi^*(u,1)\right]\,\d u=0.
\numberthis
\label{eq:statcurrents}
\end{align*}
Summing the two stationary currents, $J^*_0(\phi)+J^*_1(\phi)=0$, we immediately find $v_0(\phi)\Pi^*(\phi,0)+v_1(\phi)\Pi^*(\phi,1)=0$ for all $\phi$, i.e., $\Pi^*(\phi,1)=-\left[v_0(\phi) / v_1(\phi) \right]\Pi^*(\phi,0)$. 

Inserting this into the first equation of \eqref{eq:statcurrents} gives
\begin{align*}
&v_0(\phi)\Pi^*(\phi,0)\nonumber \\
&+\int_{\phi_0^*}^\phi
 \left[\lambda_+(u)\Pi^*(u,0)+\lambda_-(u)\frac{v_0(\phi)}{v_1(\phi)}\Pi^*(u,0)\right] \, \d u=0.
 \numberthis
\end{align*}
Differentiating with respect to $\phi$ one obtains
\begin{align*}
\rho_0'(\phi)+\left[\frac{\lambda_+(\phi)}{v_0(\phi)}+\frac{\lambda_-(\phi)}{v_1(\phi)}\right]\rho_0(\phi)=0,
\numberthis
\label{eq:help}
\end{align*}
where we have introduced $\rho_0(\phi)\equiv v_0(\phi)\Pi^*(\phi,0)$, and where $\rho_0'$ indicates a derivative with respect to $\phi$. An analogous derivation shows that $\rho_1(\phi)\equiv v_1(\phi)\Pi^*(\phi,1)$ fulfills the same relation (we note that the expression in the square brackets in Eq.~\eqref{eq:help} is symmetric with respect to simultaneous exchanges $0\leftrightarrow 1$ and $\lambda_+\leftrightarrow\lambda_-$).

Eq.~\eqref{eq:help} and its analogue for $\rho_1(\phi)$ can directly be integrated and we find
\begin{align*}
\Pi^*(\phi,0)=\frac{\mathcal{N}_0}{v_0(\phi)} h(\phi), && \Pi^*(\phi,1)=\frac{\mathcal{N}_1}{v_1(\phi)} h(\phi),
\numberthis
\label{eq:app_general_dist_phi}
\end{align*}
where 
\begin{align*}
h(\phi)= \exp \left[ - \int^\phi \d u ~ \left( \frac{\lambda_+(u)}{ v_0(u)} + \int \! \frac{\lambda_-(u)}{ v_1(u)} \right) \right],
\numberthis
\label{eq:app_general_dist_phi_h}
\end{align*}
and where $\mathcal{N}_0$ and $\mathcal{N}_1$ are normalisation constants. Using again the relation $v_0(\phi)\Pi^*(\phi,0)+v_1(\phi)\Pi^*(\phi,1)=0$, derived above, we conclude $\mathcal{N}_0=-\mathcal{N}_1\equiv -\mathcal{N}$, and so we have
\begin{align*}
\Pi^*(\phi,0) =\frac{\mathcal{N}}{-v_0(\phi)} h(\phi) , &&
\Pi^*(\phi,1)=\frac{\mathcal{N}}{v_1(\phi)} h(\phi).
\numberthis
\end{align*}
Recalling that $v_0<0$ and $v_1>0$ throughout the domain of the PDMP, $\mathcal{N}$ is a positive constant, to be determined from the normalisation condition 
\begin{align}
\int_{\phi_0^*}^{\phi_1^*} \left[\Pi^*(\phi,0)+\Pi^*(\phi,1)\right] \d\phi = 1.
\label{eq:app_general_normalisation_condition}
\end{align}
~
\begin{figure}[t]
\captionsetup[subfigure]{labelformat=empty}
{\centering
\includegraphics[width=\columnwidth]{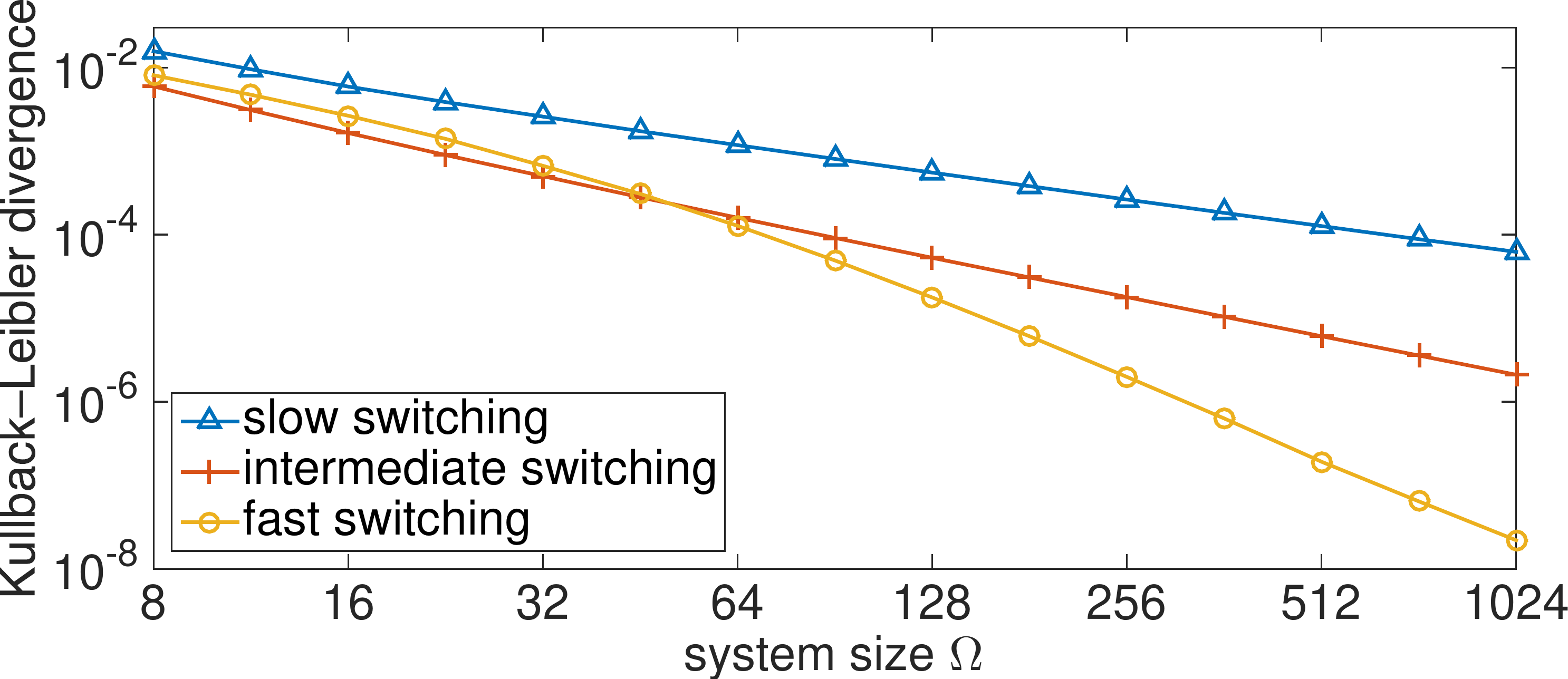}}
\caption{The Kullback-Leibler divergence between the true stationary distribution of the system and our approximation for the linear model. Each line shows a different parameter regime from Figure~\ref{fig:simple_figs_x}. The true stationary distribution is determined by fourth order Runge--Kutta integration of the master equation.}
\label{fig:KL_divergence}
\end{figure}

\section{Further details of the model with multiple fixed points}\label{app:multfp}
For the model described in Sec. \ref{sec:multfp} the functions $v_\sigma(x)$ and $w_\sigma(x)$ as defined in Eq.~\eqref{eq:general_VW} are given by
\begin{align*}
v_\sigma(x)&=
c_{1,\sigma}
-c_{2,\sigma}x
+c_{3,\sigma}x^2
-c_{3,\sigma}x^4
,\\
w_\sigma(x)&=
c_{1,\sigma}
+c_{2,\sigma}x
+c_{3,\sigma}x^2
+c_{3,\sigma}x^4.
\numberthis
\label{eq:multiple_VW}
\end{align*}
The parameters chosen for the analysis in Sec.~\ref{sec:multfp} are
\begin{align*}
c_{1,0}=0.11, && c_{1,1}=0.31,\\
c_{2,0}=0.76, && c_{2,1}=1.24,\\
c_{3,0}=2.14, && c_{3,1}=2.60,\\
c_{4,0}=2.40, && c_{4,1}=2.40.
 \numberthis
\end{align*}\\

\section{Accuracy of our approach}

The accuracy of our approach can be characterised by computing the Kullback--Leibler divergence \cite{kullback1968information} between the true stationary distribution and our approximation. Figure~\ref{fig:KL_divergence} shows how the Kullback--Leibler divergence changes with system size for the linear model. Each line represents one of the parameter regimes displayed in Figure~\ref{fig:simple_figs_x}. Birth rates, death rates, and switching rates have been chosen such that $\Omega$ is the average population of the system. The divergence decreases as the system size is increased, with relatively small system sizes having appreciably small divergences. Similar results are observed for the nonlinear and feedback models.

\end{document}